\documentclass[fleqn,usenatbib]{mnras}

\usepackage{newtxtext,newtxmath}
\usepackage[T1]{fontenc}
\usepackage{anyfontsize}
\usepackage[utf8]{inputenc}

\DeclareRobustCommand{\VAN}[3]{#2}
\let\VANthebibliography\thebibliography
\def\thebibliography{\DeclareRobustCommand{\VAN}[3]{##3}\VANthebibliography}

\usepackage{graphicx}	
\usepackage{amsmath}

\defcitealias{cusack_fragmentation_2025}{C25} 

\title[Irradiated Disc Evolution]{The Impact of Radiation Environment on the Evolution and Fragmentation of Protostellar Discs}

\author[M. T. Cusack et al.]{
Matt T. Cusack,$^{1}$\thanks{E-mail: cusackMT@cardiff.ac.uk}
Paul C. Clark,$^{1}$
Ken Rice,$^{2,3}$
Simon C. O. Glover,$^{4}$
Ralf S. Klessen,$^{4,5}$
\newauthor{
Anthony P. Whitworth,$^{1}$
Felix D. Priestley,$^{1}$}
Ana Duarte-Cabral$^{1}$
\\
$^{1}$School of Physics and Astronomy, Cardiff University, Queen's Buildings, The Parade, Cardiff CF24 3AA, UK  \\
$^{2}$SUPA, Institute for Astronomy, University of Edinburgh, The Royal Observatory, Blackford Hill, Edinburgh EH9 3HJ \\
$^{3}$Centre for Exoplanet Science, University of Edinburgh, Edinburgh EH9 3HJ \\
$^{4}$Universit\"{a}t Heidelberg, Zentrum f\"{u}r Astronomie, Institut f\"{u}r Theortische Astrophysik, Albert-Ueberle-Str. 2, 69120 Heidelberg, Germany \\
$^{5}$Universit\"{a}t Heidelberg, Interdisziplin\"{a}res Zentrum f\"{u}r Wissenschaftliches Rechnen, Im Neuenheimer Feld 225, 69120 Heidelberg, Germany \\
}

\date{Accepted XXX. Received YYY; in original form ZZZ}

\pubyear{2026}

\begin{document}
\label{firstpage}
\pagerange{\pageref{firstpage}--\pageref{lastpage}}
\maketitle

\thispagestyle{plain}


\begin{abstract}
    We present high-resolution zoom-in simulations of molecular clouds exposed to an interstellar radiation field and cosmic ray ionisation rate up to 1000 times stronger than that of the solar neighbourhood. We detail the evolution of the accretion discs that form around the first protostar in each simulation, for a total of 7 discs, for up to $\rm 100 \, kyr$. The use of a zoom-in procedure allows for the au-scale discs to be well resolved (with resolution $< 0.25 \, \rm au$) whilst retaining the structure of the wider parsec-scale molecular cloud. We find that discs exposed to a stronger radiation field tend to be more massive, hotter and denser. Similarly, their host stars grow to become more massive as a result of accreting more rapidly from their surroundings. All the discs show evidence of recurrent instability during the simulations, but only some of them fragment. We investigate whether stability metrics, such as the Toomre $Q$, $\alpha$ viscosity, and $\beta$ cooling parameter, can predict fragmentation by calculating them just before the discs fragment. We find that the metrics are generally unable to do so, as the discs appear stable even up to a few hundred years before fragmenting. In solar-like environments fragments are typically of planetary mass and often migrate to the centre of the disc, whereas fragments in a high-radiation environment are massive ($\rm > 0.1 \, M_\odot$) and fully disrupt/accrete from the progenitor disc. We conclude that the evolution and properties of circumstellar discs depend on both their radiation and physical environment.
\end{abstract}




\section{Introduction}

Accretion discs are a natural phase of the star and planet formation process, as angular momentum is conserved during the collapse of a rotating core \citep{tscharnuter_collapse_1975, boss_bipolar_1987, bodenheimer_angular_1995, machida_origin_2011}, and appear to be ubiquitous in recent surveys \citep{ansdell_alma_2016, andrews_disk_2018, cieza_ophiuchus_2019, oberg_molecules_2021, van_terwisga_survey_2022, zhang_alma_2025, teague_exoalma_2025}. Thus it is expected that discs would be present in a wide range of environments, including those with high interstellar radiation fields and cosmic ray ionisation rates. Examples of these are the nearby Orion complex and the Central Molecular Zone (CMZ), both of which have confirmed observations of circumstellar discs \citep{mann_alma_2014, stolte_circumstellar_2015, van_terwisga_survey_2022, xu_dual-band_2025}. The extreme environment presented by these locations has been shown to effect the star formation process, increasing the mass of cores and the stellar systems they create, as well as changing both the physical and thermodynamic structure of molecular clouds \citep{wolfire_neutral_2003, guszejnov_effects_2022, cusack_fragmentation_2025}. These effects will naturally impact the formation and evolution of circumstellar discs.

Circumstellar discs inherit many of their properties from their environment. The angular momentum of the parent core can affect the properties and stability of the disc \citep{forgan_lower_2012, vorobyov_effect_2015} and so too can the accretion onto the disc from its envelope, either globally \citep{bate_accretionn_1997, kratter_role_2010} or locally through streamers \citep{pineda_protostellar_2020, valdivia-mena_prodige_2022,valdivia-mena_probing_2024, calcino_anatomy_2025, speedie_mapping_2025}. The changes to core structure, mass and spatial distribution induced by strong radiation fields can therefore have a profound impact on the circumstellar discs via changes to accretion, angular momentum transport and envelope mass. Irradiation of the disc itself has also been shown to affect its evolution, either through photoevaporation \citep{adams_photoevaporation_2004, owen_theory_2012, haworth_fried_2023, keyte_impact_2025} or thermodynamic changes that impact the stability of the disc \citep{meru_exploring_2010, rice_stability_2011, forgan_effect_2013, coleman_observable_2025}. However no studies thus far have considered how the effects of irradiation on large scales cascade down to the scales of circumstellar discs.

The majority of early studies of discs either used 1/2D analytic models (e.g. those of \citealt{gammie_nonlinear_2001, vorobyov_self-regulated_2007}), or fully hydrodynamic models of isolated discs (such as \citealt{rice_substellar_2003, lodato_testing_2004, meru_exploring_2010, vorobyov_embedded_2010, rice_stability_2011}). While the physics of isolated accretion discs has been explored and well understood for many decades (see the reviews by  \citealt{pringle_accretion_1981} and \citealt{lodato_self-gravitating_2007}), the necessity to consider discs in the context of their wider environment is becoming apparent. Therefore more recently, simulations that consider disc formation beginning from cloud-scale initial conditions have become the norm and show considerable differences to analytical or isolated disc models \citep{kuffmeier_zoom-simulations_2017, bate_diversity_2018, he_massive_2023, lebreuilly_synthetic_2024, lebreuilly_influence_2024, mayer_protostellar_2025}. In particular, these simulations highlight how a more realistic star formation environment can produce a wide range of protostellar systems. The environment around the discs creates huge diversity in disc morphology, properties and evolutionary history that cannot be captured in analytic or isolated models. It is therefore vital that studies of discs in different environments begin from realistic cloud-scale initial conditions. 

The effect of the radiation environment on star formation has been studied by various authors, typically in the context of the CMZ or higher redshift environments \citep[e.g. ][]{wolfire_neutral_2003, klessen_formation_2007, chabrier_variations_2014, guszejnov_effects_2022, guszejnov_effects_2023, whitworth_minimum_2024, bate_variation_2025}. In particular, \citet[hereafter \citetalias{cusack_fragmentation_2025}]{cusack_fragmentation_2025}, studied how the strength of the ambient radiation field affects the fragmentation of molecular clouds. They found that stronger radiation fields increase gas temperatures, leading to the formation of more massive cores that in turn fragment to produce more massive stellar systems. These simulations were limited to resolution > 200 au, larger than most circumstellar discs, but the effects they describe are likely to impact the formation and growth of such discs. Therefore in this study we expand the work of \citetalias{cusack_fragmentation_2025} by performing zoom-in simulations on a number of cores in different radiation environments, resolving star formation down to the first Larson core \citep{larson_numerical_1969}. In Section \ref{sec:methods} we describe the numerical setup, including the zoom-in procedure, changes to the thermodynamic model and disc identification. In Section \ref{sec:discEvolution} we report the evolution and history of the first disc to form in each simulation, in Section \ref{sec:discEvents} we explore a series of notable events that occur in the lifetime of each disc, and in Section \ref{sec:discFragmentation} we investigate the fragmentation events that occur in the discs and their relation to stability metrics.

\section{Methodology} \label{sec:methods}

\subsection{Numerical Simulations}

In this work we re-simulate sections of the \citetalias{cusack_fragmentation_2025} clouds at much higher resolution. We again use the adaptive mesh refinement \textsc{arepo} code \citep{springel_e_2010}, including the sink particle routine and astrochemical network described in \citetalias{cusack_fragmentation_2025}. Whereas the minimum resolution element in \citetalias{cusack_fragmentation_2025} was on the order of $\rm 200 \, au$, here we follow the simulations down to sub-au scales, past the first Larson core's typical size of a few au \citep{larson_numerical_1969}. We insert sink particles above a density of $\rm 10^{-11} \, g \, cm^{-3}$, the opacity limit, with an interaction radius of $\rm 2 \, au$. The interaction radius defines the size of region that must be dense, converging and bound before forming a sink, and the radius within which a formed sink can accrete. We choose these parameters in particular as past this point in the collapse of a core, the short lifetime and rapidly increasing temperatures make it unlikely that any fragmentation will occur "inside" the sink particle \citep{bhandare_first_2018}, thus allowing us to consider the sink particles as individual stars. Otherwise, apart from some changes to the thermodynamic model outlined in Section \ref{sec:newPhysics}, the simulation setup is the same as described in \citetalias{cusack_fragmentation_2025}.

\subsection{Zoom-In Procedure}

\begin{table}
    \centering
    \caption{The properties of each of the zoom regions presented in this work. Positions indicate the centre of the zoom region and are given in Cartesian coordinates relative to the entire simulation box, including padding. Enclosed mass is the mass enclosed within the region at the beginning of the zoom simulation. Each zoom region is labelled as $\gamma_{a,b}$ where $a$ refers to the parent \citetalias{cusack_fragmentation_2025} cloud, and $b$ refers to the number of the zoom region within the cloud.}
    \begin{tabular}{c|c|c|c|c|c|c}
    \hline
    Name & $t_{\rm i}$ [kyr] & $x_{\rm i}$ [pc] & $y_{\rm i}$ [pc] & $z_{\rm i}$ [pc] & $M_{\rm Enc} \, \rm [M_\odot]$ \\
    \hline \hline

    $\gamma_{1,1}$    & 314 & 26.364 & 30.852 & 31.067 & 24.58  \\
    $\gamma_{1,2}$    & 348 & 30.144 & 32.003 & 32.564 & 65.26  \\
    $\gamma_{1,3}$    & 387 & 29.510 & 29.314 & 28.407 & 39.40  \\
    $\gamma_{1,4}$    & 408 & 31.648 & 31.165 & 32.849 & 31.90  \\
    $\gamma_{1000,1}$ & 786 & 31.320 & 30.808 & 27.343 & 182.82 \\ 
    $\gamma_{1000,2}$ & 807 & 30.188 & 32.455 & 31.495 & 114.19 \\
    $\gamma_{1000,3}$ & 943 & 30.439 & 28.918 & 29.562 & 115.10 \\

    \hline
    \end{tabular}
    \label{tab:zoomProperties}
\end{table}

The simulations we describe are zoom-in simulations of two of the clouds presented in \citetalias{cusack_fragmentation_2025}. Both clouds are turbulent and virialised with a mass of $10^4 \, \rm M_\odot$ and radius of $\rm 4.1 \, pc$. We take the solar neighbourhood case, labelled $\gamma_1$, and the most highly irradiated case, labelled $\gamma_{1000}$, and select 3-4 zoom regions in each that are just about to form a sink particle. Here, $\gamma$, or $\gamma_{\rm SFR}$ refers to the combination of the interstellar radiation field (ISRF) and cosmic ray ionisation rate (CRIR), which are taken together to be proxies for the local star formation rate. A higher star formation rate implies the production of more O/B-type stars, increasing the strength of the local ISRF, and the presence of more stellar remnants, which increases the CRIR. The solar neighbourhood case, $\gamma_1$ refers to an ISRF of $\rm 1.7$ in Habing units \citep{habing_interstellar_1968, draine_photoelectric_1978} and a CRIR of atomic hydrogen of $\rm 3.5\times10^{-16} \, s^{-1}$ \citep{indriolo_investigating_2012}. The extreme case, $\gamma_{1000}$, has an ISRF and CRIR 1,000 times as strong as this. 

Zoom regions are setup as being spherical, with radii of 0.32 pc, and chosen to include both clustered and isolated regions to better represent the range of environments present within the clouds. The properties of each region are reported in Table \ref{tab:zoomProperties}. We perform the zooms by injecting a low mass, collisionless, flow tracer particle at the location of every potential peak in the \citetalias{cusack_fragmentation_2025} simulation. These particles follow the gravitational potential of the gas around them without ever being de-refined or accreted by a sink particle. At the beginning of the re-simulation, we select the tracer particle closest to the centre of the desired zoom region and define that particle as the centre of the region. The tracer acts as an "anchor"; moving the centre of the zoom region with the core as it moves through the wider cloud, keeping it continually refined.

Inside the zoom region we refine cells according to a Jeans refinement criterion. Cells will refine or de-refine such that the local Jeans length is always resolved with at least 16 cells, until the sink creation density is reached. This gives a maximal spatial resolution of $\rm 0.23 \, au$ inside the zoom region, assuming a temperature of $\rm 200\, K$ at the sink creation density (motivated by the $\gamma_1$ case in Figure \ref{fig:beTempDense}). This resolution scales continuously with density and temperature, as \textsc{arepo} does not rely on discrete levels of refinement as many grid codes do and can instead smoothly transition between regions of low and high density. Additionally, cell smoothing lengths are similarly continuous and depend on the size of the cell, ensuring that gravitationally relevant features are always resolved.

Outside the region, we refine according to a target mass condition, such that cells will merge or split in order to attain a desired mass. This mass increases with radius from the zoom, ensuring that the resolution decreases smoothly with distance from the region:

\begin{equation}
    M_{\rm Target} (R) = M_{\rm Resolution} + 0.2 \, {M_\odot} \frac{(R_{\rm Centre} \, \rm / \, pc)^{1.3}}{(L_{\rm Box}\,/\rm \,pc)}
\end{equation}

\noindent where $M_{\rm Target}$ is the desired mass at that radius, $M_{\rm Resolution}$ is the initial target mass of the \citetalias{cusack_fragmentation_2025} simulation ($\rm 0.01 \, M_\odot$), 0.2 is a scaling factor, $L_{\rm Box}$ is the side length of the simulation box (62 pc), and $R_{\rm Centre}$ is the cell's distance to the centre of the zoom region. This gradual de-refinement prevents shocks from occurring at the boundary of the zoom region, whilst reducing the overall number of cells in the simulation. 

We retain the cells in the wider cloud to maintain the gravitational potential map around the zoom region. Therefore, we are not interested in their chemical evolution and can stop evolving their chemical state. The heating and cooling rates of the ISM are ignored, and the cells act as an ideal gas with a polytropic index of 5/3. The only changes to the thermal state of the cells is due to hydrodynamic work and shocks. We do not expect this to affect the gas within the zoom, as the size of the regions is chosen such that any infall from the surrounding cloud is a small compared the mass already contained within them. We also do not expect the zoom region to influence the chemical evolution of the wider cloud due to the lack of radiative feedback, nor do we expect the chemical state of the gas outside the region evolve sufficiently to become dynamically important on the timescales we are investigating.

\subsection{Thermodynamical Changes} \label{sec:newPhysics}

Accurately modelling the gas at densities beyond those in \citetalias{cusack_fragmentation_2025} requires the consideration of new physics that is not present in the \citetalias{cusack_fragmentation_2025} simulations. We make two notable changes to the thermodynamic model: dust cooling now becomes optically thick to its own emission at high densities, and the adiabatic index of molecular hydrogen is no longer constant, but changes as its temperature increases \citep{black_evolution_1975}. 

At high densities ($\rm n > 10^5 \, cm^{-3}$) dust cooling is central to the thermal balance of the ISM. However, as cores collapse the emission from dust will eventually become optically thick and no longer cool the gas effectively. To take this into account, we implement a modified version of the dust cooling from \citet{stamatellos_radiative_2007}. In this model the cooling rate per unit volume from dust is given by 

\begin{equation}
    \Gamma_{\rm dust} = \rho \frac{4 \sigma_{SB} T_{\rm dust}^4}{\alpha_{\rm dust}}, \,\rm with
    \label{eq:dustEmission}
\end{equation}

\begin{equation}
    \alpha_{\rm dust} =
    \begin{cases}
        a \,\Sigma^2 \, \left(\frac{T_{\rm dust}}{\rm K}\right)^2 + \frac{1}{a \, \left(\frac{T_{\rm dust}}{\rm K}\right)^2} \quad \text{for $T_{\rm dust} \le 150 \, \rm K$} \\
        b \, \Sigma^2 \, \left(\frac{T_{\rm dust}}{\rm K}\right)^{1/2} + \frac{1}{b \, \left(\frac{T_{\rm dust}}{\rm K}\right)^{1/2}} \quad \text{for $T_{\rm dust} > 150 \, \rm K$}
    \end{cases}
\end{equation}

\noindent where $\sigma_{SB}$ is the Stefan-Boltzmann constant, $\rho$ is the density of the gas, $T_{\rm dust}$ is the temperature of the dust and $\Sigma$ is the column density of hydrogen. Both $a$ and $b$ are constants with values $2\times10^{-4} \, \rm cm^{2} \, g^{-1}$ and $0.367 \, \rm cm^{2}\,g^{-1}$ respectively. The difference between the sub and super $150 \, \rm K$ components of $\alpha_{\rm dust}$ correspond to the changing opacity coefficient of the dust as ice grains evaporate \citep{stamatellos_radiative_2007}. The opacity laws are derived from \citet{bell_using_1994}, but with a modified coefficient for the $> \rm 150 \, K$ case. This coefficient is modified so that the transition between the two cases of $\alpha_{\rm dust}$ is continuous, which is required for the iterative dust temperature solver. This approach also differs from \citet{stamatellos_radiative_2007} as the column density is calculated from a weighted average of the column densities as seen by the gas cell in each direction, given by \textsc{treecol} \citep{clark_treecol_2012}, rather than an approximate pseudo mean column density. The cooling rate, $\Gamma_{\rm dust}$, along with the heating terms from the ISRF, is then used to determine the equilibrium temperature of the dust (independently of the gas temperature) and to calculate the gas-grain energy transfer. The effect of this change is that above densities of $\approx 10^8 \, \rm cm^{-3}$, gas temperatures tend to rise with increasing density. 

We assume that the gas and dust are dynamically coupled and well mixed, with a constant dust-to-gas ratio of 0.01 and a fixed MRN grain size distribution \citep{mathis_size_1977}. This distribution assumes maximum grain sizes of approx $\rm 0.3 \, \mu m$, below the threshold for which dust-gas dynamical decoupling becomes important ($> 4 \rm \,\mu m$, \citealt{commercon_dynamics_2023}). At disc midplane densities coagulation may produce a population of large, micron-sized grains from the initial MRN distribution on the timescales we consider here \citep{bate_modelling_2026}, which would not be fully coupled to the gas dynamics. This would lead to variations in the dust-to-gas ratio throughout the disc, potentially affecting the thermal structure through dust-linked heating and cooling processes. However, detailed modelling of the dust spatial and size distribution is beyond the scope of this paper, which is primarily concerned with how interaction with the external cloud material affects the gas dynamics of protostellar discs

At temperatures above $\rm 100\,K$, the rotational modes of molecular hydrogen become excited, changing its effective adiabatic index from $5/3$ to $7/5$. This reduces the rate of temperature increase with density and is dynamically important to the formation of the first Larson core \citep{bhandare_first_2018}. To account for this, we change the adiabatic index of the gas depending on its temperature and $\rm H_2$ fraction. \textsc{arepo} provides native support for this through the HLLD Riemann solver\footnote{This solver is typically used for, and supports, magnetohydrodynamical simulations. However we do not consider magnetic fields in this work.} \citep{pakmor_magnetohydrodynamics_2011}, which calculates the adiabatic index for a given temperature and molecular mix using the methods of \citet{boley_internal_2007}.

\begin{figure}
    \centering
    \includegraphics[width=\columnwidth]{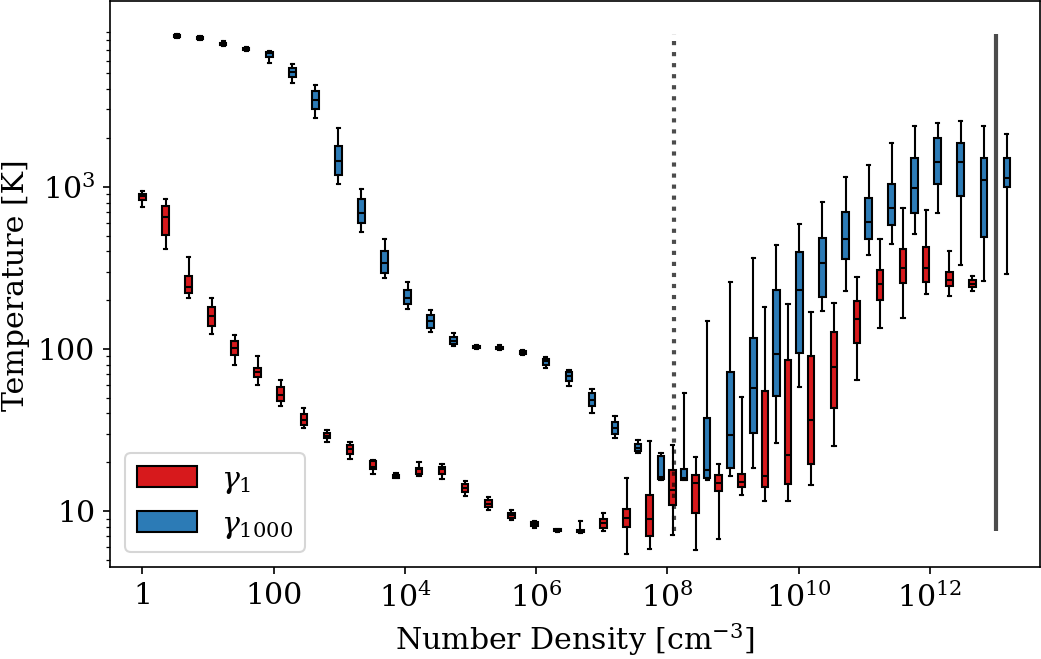}
    \caption{The relationship between gas temperature and the number density of hydrogen for the collapse of a Bonnor-Ebert sphere with a mass of $\rm 1 \, M_\odot$, for the $\gamma_1$ and $\gamma_{1000}$ models. The dotted black line denotes the density above which \citetalias{cusack_fragmentation_2025} inserted sink particles, while the solid line denotes where sink particles are inserted in this study. These act effectively as an upper limit on the densities achieved in the respective simulations. Below the sink creation density of \citetalias{cusack_fragmentation_2025}, both sets of thermodynamic models behave identically. The box plot whiskers show the 5-95\% range of temperatures for that density bin.}
    \label{fig:beTempDense}
\end{figure}

The result of both changes are evidenced in Figure \ref{fig:beTempDense}. The new model displays rapid heating as the density rises above $\rm 10^8 \, cm^{-3}$, due to dust emission becoming optically thick. Both the $\gamma_1$ and $\gamma_{1000}$ models display this behaviour, though $\gamma_{1000}$ begins this heating slightly earlier. The $\gamma_1$ model also displays a "turn over" above $\rm 10^{11} \, cm^{-3}$, where the rise in temperature slows as a result of the change in the adiabatic index of hydrogen. This is seen to a lesser degree in the $\gamma_{1000}$ model, and occurs at a higher density.

\subsection{Disc Identification} \label{sec:discIdentification}

Identifying cells that are members of a disc is a complex and sometimes arbitrary process. Studies in this field have taken various approaches, from primarily considering velocities \citep{joos_protostellar_2012}, orbital eccentricity \citep{bate_diversity_2018}, to density \citep{mayer_protostellar_2025}. For this work we have developed an algorithm, referenced as \textsc{discfind}, which primarily follows \citet{bate_diversity_2018} in that it considers the orbital parameters of cells to determine disc membership. The algorithm undertakes the following steps to identify discs in the simulation: 

\begin{enumerate}
    \item Loop through the gas cells and determine which sink each is most gravitationally bound to, i.e which sink-cell pair has the lowest binding energy.
    \item Using the cells that are most bound to it, determine the members of each sink's circumstellar disc. This is done with the following steps:
    \begin{enumerate}
        \item Sort the candidate gas cells by distance to the sink, closest first. Loop through the cells in distance order.
        \item Determine the instantaneous ballistic orbit of the cell around the sink.
        \item Add the cell to the disc if its orbital eccentricity is $< 0.5$ and semi-major axis is $< 2,000 \, \rm au$. 
        \item Update the centre of mass and velocity of the disc if the cell is within $50 \, \rm au$ of the sink.
        \item Terminate if the cell is more than $2,000 \rm \, au$ from the sink, or if another sink particle is encountered. Sink particles of a very low mass ($< 0.05 \, \rm M_\odot$) are ignored in this step as they are generally fragments that have migrated to the centre of a disc alongside the host star.
    \end{enumerate}
    \item Loop through the gas cells and determine which binary system (if any are present) each are most bound to.
    \item Using the cells that are most bound to it, determine the members of each circumbinary disc. This is done using the same method as for the circumstellar disc.
    \item Repeat steps (i)-(iv), now including the found disc masses in addition to the sink or binary masses when determining which system gas cells are most bound to. 
\end{enumerate}

Once the discs have been determined using the algorithm, they are truncated based on a surface density cut. Discs found using the orbital parameter method can often have extended or disconnected cells with low surface density. These regions are generally not part of the disc itself, but just happened to meet the criteria for disc membership. To account for this we calculate a radial profile of surface density for each of the discs, and truncate the disc at the radius where its surface density drops below $5 \, \rm g \, cm^{-2}$. This value is chosen as when the surface-density profiles of the discs fall below it they become especially noisy, and it excludes much of the extraneous envelope material. For circumbinary discs we ensure that the disc is not truncated inside the binary semi-major axis, as circumbinary material is by definition material outside this radius. Finally, any discs with a mass $< 0.03 \, \rm M_\odot$ or radius $< 15 \rm \, au$ are rejected as discs below this size are generally not well resolved and this is close to the size of the first optically thick core \citep{bhandare_first_2018}. The sensitivity of the mass and radius of the discs to the choice of the eccentricity and surface density cuts is discussed in Appendix \ref{sec:discParameters}.

\section{Disc Histories} \label{sec:discEvolution}

Each zoom simulation was evolved for up to 200 kyr. This allows enough time for the cores to collapse and form stars, but not enough for photoevaporation-induced mass loss to significantly deplete the discs \citep{haworth_fried_2023} or for planet formation mechanisms to become important. Some of the simulations, $\gamma_{1,2}$ and $\gamma_{1,3}$, were terminated before this point due to computational constraints. An error in the snapshot output rate caused the 75-99 kyr evolution of the $\gamma_{1000,1}$ disc to be skipped, leading to a gap it its evolution but retention of the 100 kyr panel in subsequent Figures. 

Many sink particles form in each zoom region, but in this study we follow only the first sink to form in each simulation. This is because these sinks have the longest lifetimes, however they can still take many tens of kyr to form after the simulations begin. For this reason the longest common lifetime between all the sinks in our sample is 100 kyr, and we follow their evolution for this period only.

\subsection{Disc Property and Morphology Evolution} \label{sec:evolve}

\begin{figure*}
    \centering
    \includegraphics[width=0.87\textwidth]{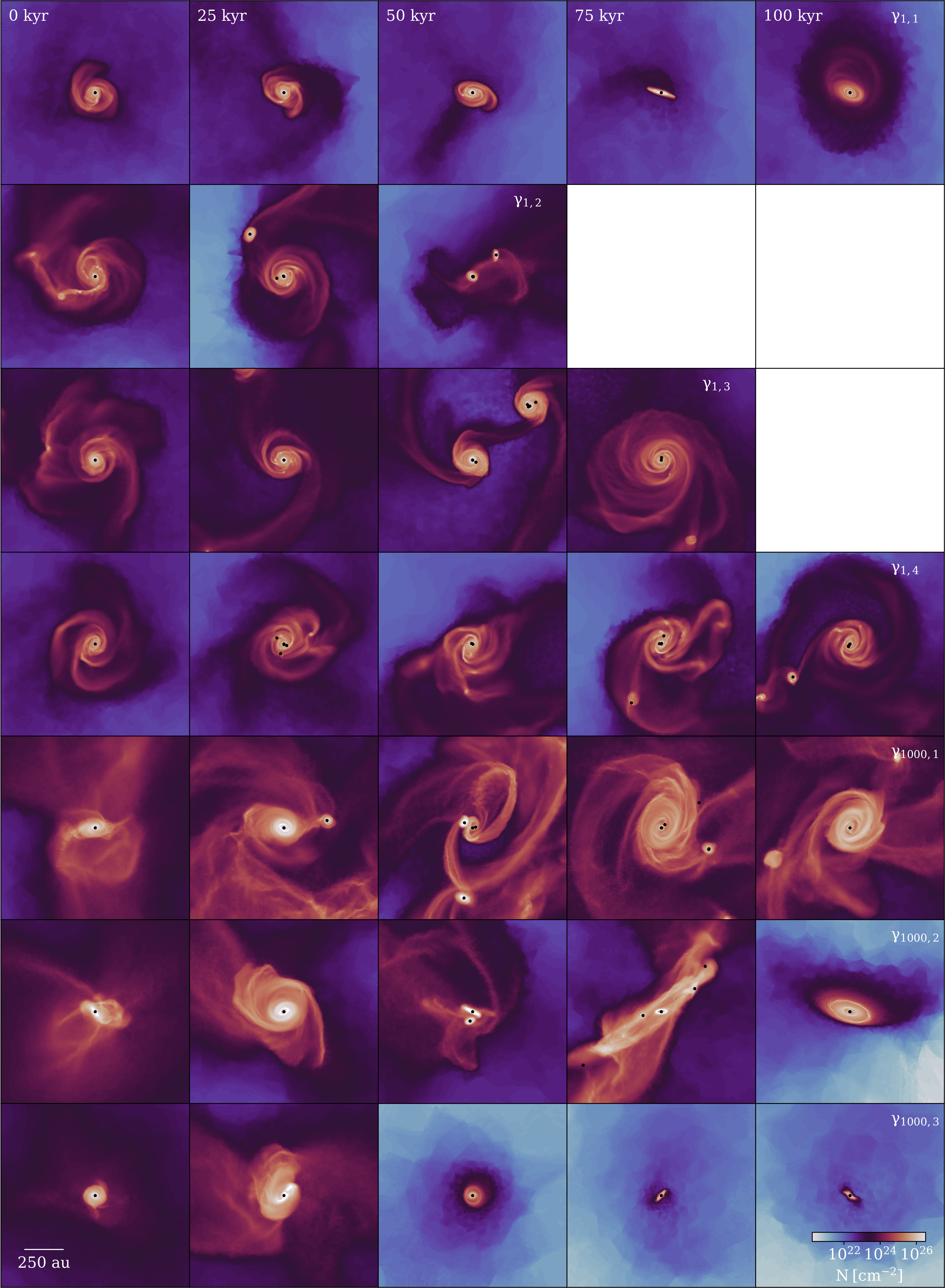}
    \caption{Column density maps of the first disc to form in each zoom simulation, shown in the x-z plane of the simulation box. Each row shows the evolution of a particular disc at intervals of $\rm 25 \, kyr$, beginning at the point of sink formation (labelled 0 kyr). The time since sink formation is shown explicitly for the top row, and the final column shows the zoom simulation label for each disc. Each panel shows a region 1200 au across and the location of sink particles are denoted with black dots. An animated version of the evolution of each disc are available in the supplementary online materials. Blank panels in the $\gamma_{1,2}$ and $\gamma_{1,3}$ cases are the result of needing to stop the runs prematurely due to computational constraints. Images of the whole zoom region, rather than the individual discs, are shown in Appendix \ref{sec:zoomPanelImage}.}
    \label{fig:firstDiscsPanel}
\end{figure*}

\begin{figure*}
    \centering
    \includegraphics[width=0.98\textwidth]{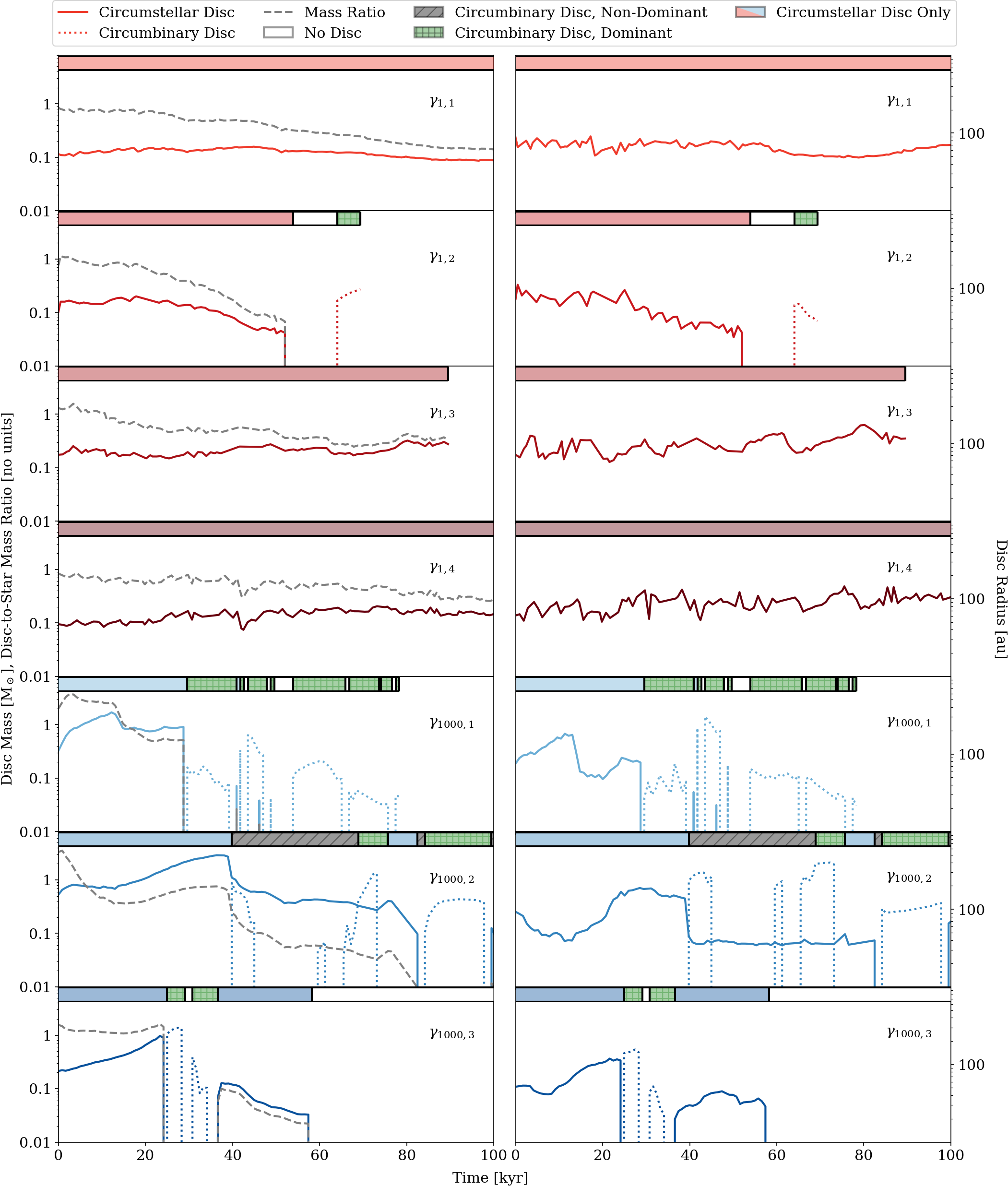}
    \caption{The evolution of the mass and disc-to-star mass ratio (left-hand panels) and radius (right-hand panels) of the
    discs around the first sink to form in each simulation, sampled every 1 kyr. The properties of the sink's circumstellar disc are plotted with a solid line, while the properties of any circumbinary disc around the sink is dotted. The coloured bars at the top of each panel indicate whether the star is part of a multiple system, and whether the circumbinary disc around that system is more massive than the star's circumstellar disc. The disc-to-star mass ratio is denoted by a dashed line, and is always calculated using the circumstellar disc mass, which may cause the line to disappear when the disc is part of a binary and no circumstellar disc exists. Breaks in the evolution occur when no disc is detected around the star. Simulations derived from the $\gamma_1$ cloud are shown in shades of red, while those derived from $\gamma_{1000}$ are shown in shades of blue.}
    \label{fig:firstDiscsEvolution}
\end{figure*}

Figures \ref{fig:firstDiscsPanel} to \ref{fig:firstDiscsEvolution} detail the histories of the first discs. In Figure \ref{fig:firstDiscsPanel} we show the morphological evolution of the discs, and in Figure \ref{fig:firstDiscsEvolution} we illustrate the evolution of their properties. Together they outline how the discs in each simulation evolve through time, including changes such as fragmentation into a binary system or destruction of the discs due to some disruptive event.  

The disc in $\gamma_{1,1}$ exhibits the simplest evolutionary history of the discs. It is completely isolated, and maintains a steady mass and radius across its lifetime. The disc has achieved a steady state, during which mild gravitational instability (GI) regulates its stability \citep{lodato_self-gravitating_2007}. Despite being comparable in mass to its host star the disc does not show evidence of large spiral perturbations, implying that it maintains local angular momentum transport, the necessary condition for a steady state \citep{lodato_testing_2004, lodato_testing_2005, forgan_nature_2011}. There are some episodes where non-axisymmetric low wavenumber (otherwise known as low-$m$ modes) perturbations arise in the disc, as seen in both the 25 and 50 kyr panels, but this does not result in fragmentation or disruption. Finally, the disc changes its orientation through time, and ends the simulation with a misaligned inner and outer disc, as a result of it accreting material with a different angular momentum vector to what is already present in the disc (explored in detail in Section \ref{sec:misalignment}). 

The discs in $\gamma_{1,2}$ through $\gamma_{1,4}$ experience stronger episodes of GI during their lifetimes. While the radii and masses of the discs are comparable to $\gamma_{1,1}$, they all produce spiral arms that extend beyond 100 au. These arms form as a consequence of the disc redistributing angular momentum, which can occur during accretion of highly polarised material, such as a streamer \citep{calcino_anatomy_2025}, or when the disc-to-star mass ratio is high \citep{lodato_testing_2005}. Inside the discs many dense clumps can be seen forming, as a result of interacting high wavenumber non-axisymmetric perturbations (high-$m$ spiral modes, \citealt{xu_globall_2025}). These clumps, however, are generally transient structures and do not persist beyond a few kyr. While still a form of fragmentation, we only consider parcels of gas that eventually form a sink particle to be true "fragments". These transient clumps are generally unassociated with the gas that forms such fragments, distinguishing them as a separate phenomenon (see Section \ref{sec:discFragmentation} for details on the origin of fragments). 

The $\gamma_{1,2}$ disc undergoes repeat interactions with another disc after 25 kyr, disrupting and then truncating it before the pair form a circumbinary disc. This disruption is unexpected for interactions between low ($\rm 0.1 \, M_\odot$) mass discs, as these typically do not cause disruption or fragmentation \citep{lodato_role_2007, forgan_stellar_2009}. However, while low in absolute mass this disc has a high disc-to-star mass ratio, for which encounters have a larger impact \citep{cadman_binary_2022}, potentially explaining the disruption. The $\gamma_{1,3}$ disc also undergoes interactions with another disc, though these encounters do not noticeably affect it. The radii of both discs do show significant variation, however, which is a common consequence of disc-disc interactions \citep{rosotti_protoplanetary_2014, munoz_stellar_2015, bhandare_effects_2016}. 

Despite similar masses, mass ratio, radii and radiation environment, the $\gamma_1$ discs all experience different evolutionary histories. The diverse physical environments around the discs leads to variations in infall and interaction history that shape their evolution. Infall in particular can determine the level of GI in a disc \citep{harsono_global_2011, kuffmeier_episodic_2018}, its structure \citep{vorobyov_effect_2015, kuffmeier_misaligned_2021, terebey_dynamics_2025, calcino_anatomy_2025}, and whether it fragments \citep{kratter_role_2010}. The variation in this infall between the discs is likely a primary driver for their differing evolutions.

The discs in $\gamma_{1000}$ experience a very different environment to their $\gamma_1$ counterparts. Increasing $\gamma_{\rm SFR}$ leads to higher gas temperatures and increased core masses (as noted in \citetalias{cusack_fragmentation_2025}), producing a denser and more chaotic environment. This is immediately evident from Figure \ref{fig:firstDiscsPanel}, where the $\gamma_{1000}$ discs are immersed in gas with a higher column density and can be seen to undergo more disruptive interactions.\footnote{Appendix \ref{fig:zoomPanel} highlights the wider environment of both sets of simulations, highlighting the physical differences between the $\gamma_1$ and $\gamma_{1000}$ cores.}

The $\gamma_{1000}$ discs initially all grow to become massive, on the order of $1 \rm \, M_\odot$. At some point they all undergo fragmentation, forming a stellar binary that destroys or significantly shrinks the original disc. Following fragmentation the discs remain disrupted for a significant length of time, > 25 kyr. Small ring-like discs form around the original protostar and its companion, surrounded by an envelope of circumbinary material leftover from the original massive disc. Sometimes the binaries harden and the disc does re-form, as it does in $\gamma_{1000,1}$, however even in this case the new disc never achieves its previous mass. 

Finally, all three $\gamma_{1000}$ first sink are ejected from their natal core by the end of the simulation. This is despite them forming first in the cores, and their central stars having attained significant masses by the time of ejection. Because the cores in $\gamma_{1000}$ are so massive, they fragment to produce clusters of sinks that quickly grow beyond $1 \rm \, M_\odot$. These clusters of massive stars undergo many N-body interactions, ejecting the first stars in the process. Following ejection the discs around the first stars are retained, but they generally shrink and become undetectable as they are accreted without replenishment from their core. 

It is evident from the $\gamma_{1000}$ discs that the ambient radiation field has a significant impact on the properties and evolution of the discs, through its ability to affect the physical environment around them. The density around the discs affects the infall rate onto them and the frequency of interactions between discs. Both of these factors are of primary importance for how a disc evolves, for discs in both low and high-$\gamma_{\rm SFR}$ environments. 

\subsection{Sink History} \label{sec:sinks}

The evolution of a circumstellar disc is naturally tied to that of its central star. Discs are reservoirs of gas from which stars accrete, and any gas accreted by a star must first travel through the disc. Through the action of viscosity (driven by turbulence, gravitational torques or magnetic effects) discs transport angular momentum outwards, allowing the innermost gas to be accreted. The accretion rate of the central star is thus linked to the properties of the disc, and provides information as to how rapid this angular momentum transport is. 

\begin{figure}
    \centering
    \includegraphics[width=\columnwidth]{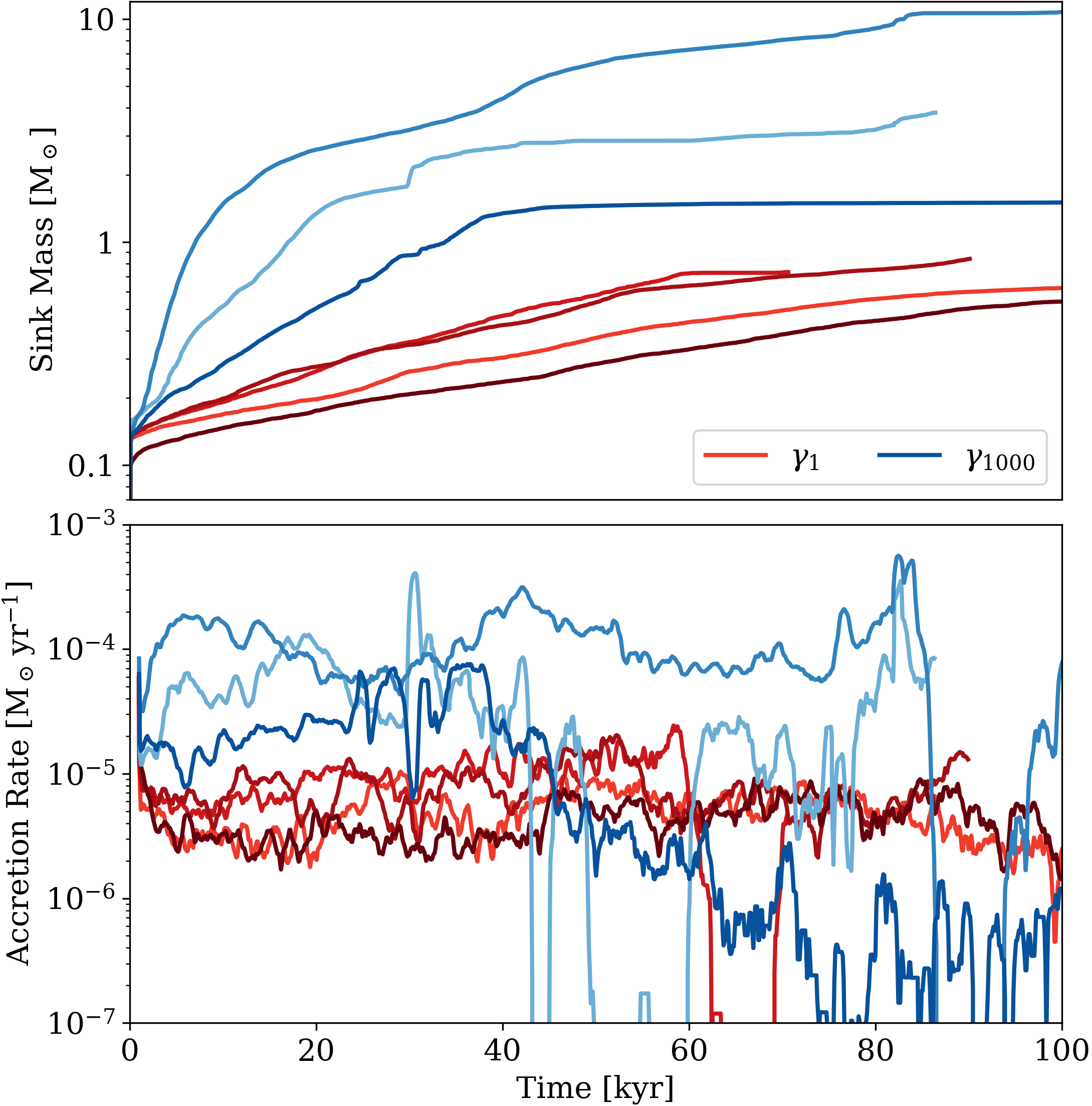}
    \caption{The mass and accretion rates of the first sinks to form in each simulation across time. Sink masses are updated every 100 years, but contain some numerical noise due to the resolution around the sink particles. Therefore the plotted accretion rate is a rolling average calculated every 1 kyr. Simulations derived from the $\gamma_1$ cloud are shown in shades of red, and those derived from $\gamma_{1000}$ in shades of blue.}
    \label{fig:sinkMassHistory}
\end{figure}

In Figure \ref{fig:sinkMassHistory} we show the evolution of the mass and accretion rate for the first sinks. In confirmation of the results of \citetalias{cusack_fragmentation_2025}, sink masses and accretion rates are higher in the $\gamma_{1000}$ clouds. One of the $\gamma_{1000}$ zooms has produced a sink with a mass exceeding $10 \, \rm M_\odot$ in under 80 kyr, a unique result for simulations of this resolution. This is due to the very high accretion rates of the $\gamma_{1000}$ sinks, typically on the order of $10^{-4} \,\rm M_\odot \, yr^{-1}$. The $\gamma_1$ sinks typically accrete at a rate an order of magnitude lower than this. The slower accretion rate onto the $\gamma_1$ sinks implies similarly slow accretion through their discs. This may in part explain why they fragment less disruptively, as often it is when discs cannot adequately redistribute the angular momentum of the material they accrete that they become unstable \citep{kuffmeier_episodic_2018}

Both sets of simulations show variability in the accretion rates of the stars. This variability is mild for the $\gamma_1$ sinks, typically within a factor of a few, and much stronger for the $\gamma_{1000}$ sinks. Following the ejection of the sinks from the core, accretion onto the sinks slows or stops completely. This implies that star formation may be rapid in high-$\gamma_{\rm SFR}$ environments, as the main accretion phase is over for all the sinks in less than 100 kyr. 

The short-lived accretion phases of the $\gamma_{1000}$ sinks suggests that protostars in high-$\gamma_{\rm SFR}$ environments may experience correspondingly shorter Class 0 phases. By 100 kyr the $\gamma_{1000}$ sinks are generally accreting at rates orders of magnitude less than they were at the start of the simulation, and are no longer embedded in their natal cores. This indicates that they have left the main accretion phase, and would be detected as Class I/II protostars. This contrasts with the $\gamma_1$ sinks, which are all still embedded with their core and accreting at rates comparable to the beginning of the simulation. These would likely still be detected as Class 0 protostars after 100 kyr.

\subsection{Global Disc Stability Evolution} \label{sec:toomre}

Instabilities, gravitational or otherwise, can drive angular momentum transport through a disc by driving turbulent viscosity. This viscosity is generally parametrised according to \citet{shakura_black_1973} and \citet{gammie_nonlinear_2001}, where the accretion rate through the disc can be used to calculate its viscosity,

\begin{equation}
    \alpha = \frac{G \bar{v}_r \Sigma \pi R}{c_s^3 \left| \frac{\rm d \ln \Omega}{\rm d \ln R}\right|} \,,
\end{equation}

\noindent where $c_s$ is the sound speed, $\bar{v}_r$ is the radial velocity through the disc, $R$ is the radius, and $\Omega(R)$ is the angular speed as a function of radius which is used to then calculate $\frac{\rm d \ln \Omega}{\rm d \ln R}$. For the discs in our sample we assume that they obey Keplerian rotation (which is generally a good assumption when the discs are unperturbed, see Appendix \ref{sec:kep}), in which case $\frac{\rm d \ln \Omega}{\rm d \ln R} = -\frac{3}{2}$. The typical upper limit of $\alpha$ for a disc before undergoing gravitational fragmentation varies between studies, and can be larger for discs under irradiation \citep{kratter_fragment_2011, rice_stability_2011}, but is typically small and on the order of 0.06 \citep{rice_investigating_2005}. Discs with more intense GI and rapid angular momentum transport will experience higher viscosity, allowing $\alpha$ to be used as a measure of instability. 

The presence of GI in a disc can also be quantified by the Toomre parameter \citep{toomre_gravitational_1964}, which measures the ratio between susceptibility to gravitational fragmentation and support by pressure and rotational forces. It is denoted as $Q$,

\begin{equation}
    Q = \frac{c_s \kappa}{\pi G \Sigma} \, ,
    \label{eq:toomre}
\end{equation}

\noindent where $\kappa$ the epicyclic frequency (which we assume is equal to $\Omega$, as is the case under Keplerian rotation) and $\Sigma$ the surface density. For thin discs instability sets in at $Q$ < 1, due to the growth of axisymmetric perturbations. Discs with realistic thickness can be subject to non-axisymmetric perturbations, which instead grow when $Q$ $\le$ 1.5 \citep{durisen_protostars_2007}. Almost all the discs in Figure \ref{fig:firstDiscsPanel} show evidence of non-axisymmetric (spiral-like) disturbances, implying $Q$ $\le$1.5 for the discs at all times.

\begin{figure*}
    \centering
    \includegraphics[width=\textwidth]{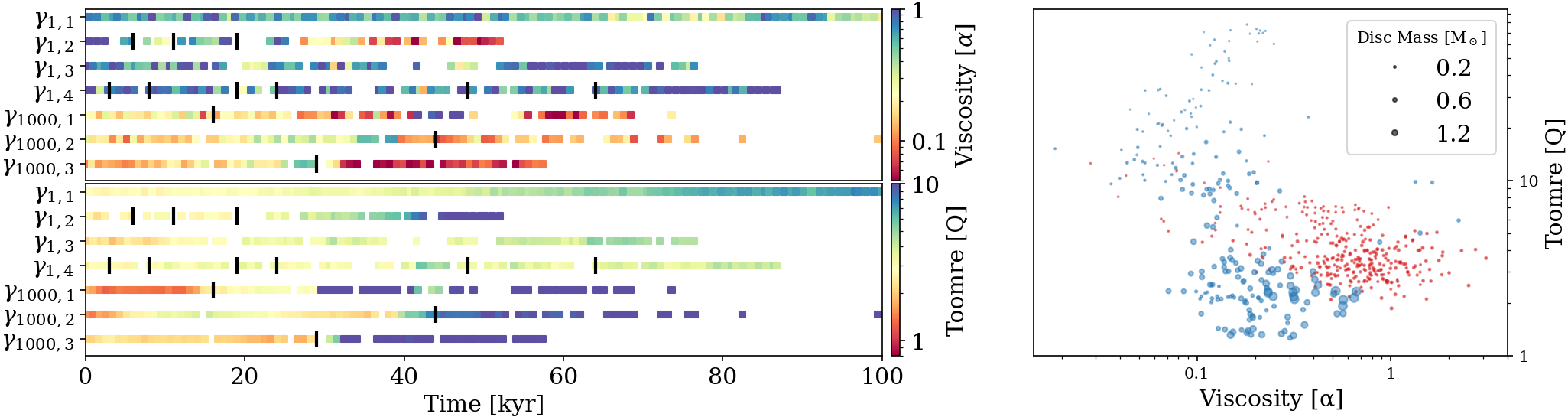}
    \caption{The viscosity and Toomre stability parameters of the first disc to form in each zoom simulation, sampled every 1 kyr. The left-hand panel shows the temporal evolution of the parameters for each disc while the right-hand panel shows the correlation between the two parameters, with points scaled in size based on the mass of the disc at that point in time. The parameters are calculated for the circumstellar disc around the first sink only, and gaps in the temporal evolution are due to periods where no disc was detected. Fragmentation events in the disc are denoted with vertical black lines. Colours for the right-hand panel denote the different parent clouds, as in previous Figures.}
    \label{fig:discViscosityEvolution}
\end{figure*}

In Figure \ref{fig:discViscosityEvolution} we illustrate the evolution of, and relationship between, the $Q$ and $\alpha$ parameters for the discs. These values are global mass-weighted averages of $Q$ and $\alpha$ across the whole disc. The disc is first subdivided into annuli of radius 10 au and an average value of $Q$ and $\alpha$ is determined for each annulus. The global average value for the whole disc is then calculated by taking an average of the values of all annuli, weighted by the mass in each annulus. 

All the discs appear Toomre-stable across the entire simulation, with $Q$ typically above 2. This is a little higher than expected, as typically discs will self regulate to the point of marginal Toomre stability via the development of gravitational instabilities that dissipate energy throughout the disc, stabilising them \citep{lodato_testing_2004, lodato_role_2007, kratter_gravitational_2016}. At later times $Q$ is large for many of the discs, as a result of truncating interactions that leave only the highly stable inner disc \citep{whitworth_minimum_2006, stamatellos_can_2008, vorobyov_formation_2010, forgan_jeans_2011}. Despite generally displaying $Q$ values greater than 1.5, many of the discs experience the recurrent growth of spiral perturbations. This is in tension with the minimum value quoted above, suggesting that the critical value of $Q$ for the growth of such perturbations may exceed 1.5 in some cases.

The $\alpha$ viscosity of the discs is generally on the order of unity for the $\gamma_1$ discs, and on the order of $0.1$ for the $\gamma_{1000}$ discs. These values far exceed the previously quoted limit for fragmentation of 0.06. Additionally the rapid accretion onto the $\gamma_{1000}$ discs and their higher mass would imply similarly rapid angular momentum re-distribution, necessitating a larger viscosity. Yet the viscosity in the slowly accreting, generally more quiescent, $\gamma_1$ discs is significantly larger. This indicates that the $\alpha$ parametrisation may be inadequately quantifying the process of angular momentum transport in our discs.

The right hand panel of Figure \ref{fig:discViscosityEvolution} showcases the relationship between the $\alpha$ and $Q$ parameters of the discs. They are moderately negatively correlated\footnote{With a Spearman's rank correlation  coefficient of 0.52 at a sub 5\% p-value.}, which is expected for discs in a self-regulated state \citep{bertin_class_1999}. The imperfect correlation indicates that the discs often experience departures from a steady state, or that the $Q$ and $\alpha$ values are not accurately describing the state of the discs. 

Fragmentation events do not generally succeed, or are followed by, systematic changes in either the $Q$ or $\alpha$ parameters. Only in the case that the disc is significantly disrupted by the fragmentation event is there a noticeable change in either parameter. Discs that do not undergo such catastrophic disruptions do not show any appreciable change in $Q$ or $\alpha$. This implies that the disc-averaged values of these parameters are not affected by, or predict, fragmentation events. 

In this Section we have noted some issues with both the $Q$ and $\alpha$ parameters for these discs. However it may not be the case that these parameters themselves have issues, but rather that they cannot realistically be applied to discs we simulate. The Toomre parameter, being a stability criterion, is only defined for an equilibrium disc and therefore will not be able to adequately describe a constantly evolving disc with time-varying properties and environment. Additionally, $\alpha$ relies on the thin-disc approximation, which is generally invalid for these discs, and has been found to inadequately describe discs with a high disc-to-star mass ratio \citep{vorobyov_applicability_2010}. Similarly, in calculating both $Q$ and $\alpha$ we have assumed perfect Keplerian rotation. While this is generally a good assumption for the discs when they are unperturbed and isolated, it quickly breaks down in the presence of external environmental effects. Interacting discs (explored more in Sections \ref{sec:discInteraction1} and \ref{sec:discInteraction2}) can exert tidal forces on each other, producing warps and toques that alter their rotation curves. The accretion of misaligned material onto the discs can also affect their structure and velocity distribution by producing an inclination difference between the inner and outer disc (explored in Section \ref{sec:misalignment}). These external factors significantly complicate the calculation of these stability metrics, and weaken their applicability to disc fragmentation.

\subsection{Disc Radial Profile Evolution} \label{sec:radial}

While the changed environment due to a higher $\gamma_{\rm SFR}$ does not produce a systematic change in the Toomre-stability of the discs, it will increase their temperature as the heating from the ISRF and cosmic rays is intensified. This can be quantified with a radial profile, showing explicitly how the discs' temperatures vary across their extent. The environment around the discs in the high-$\gamma_{\rm SFR}$ clouds is generally denser too, as we noted in Figure \ref{fig:firstDiscsPanel}. This will affect their surface density, which can also be quantified using a radial profile. However, the discs are constantly evolving structures and a radial profile at one point in time does not capture this evolution. Instead, we calculate the distribution of the radial profiles of the discs; i.e the range of temperature and surface density experienced by the disc at a given radius across its entire lifetime. 

\begin{figure*}
    \centering
    \includegraphics[width=\textwidth]{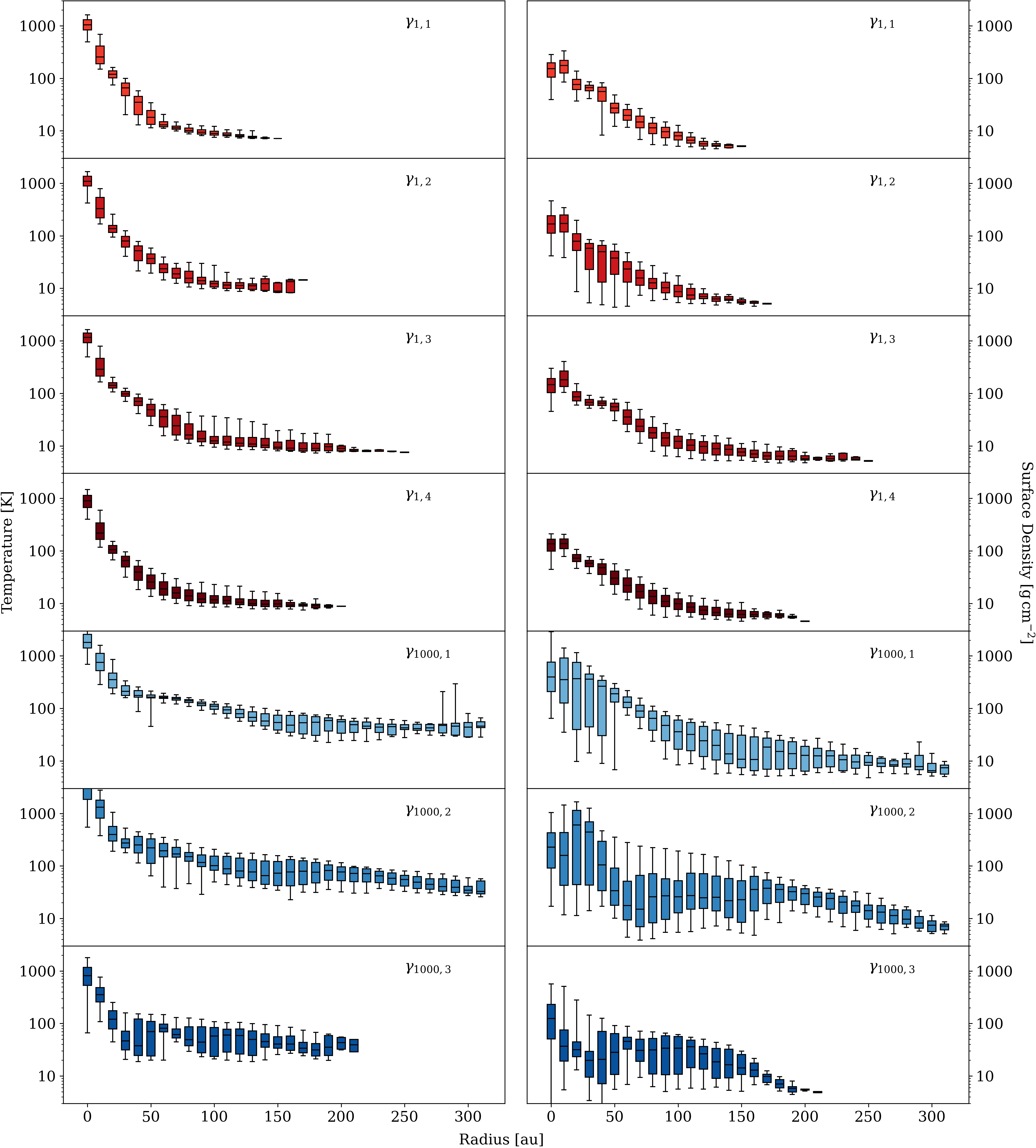}
    \caption{The radial profiles of gas temperature (left-hand panels) and surface density (right-hand panels) for each of the discs, aggregated across time. Temperature and density radial profiles are calculated for each disc every 1 kyr (the minimum temporal resolution). These profiles are then aggregated, such that the box plot at each radius shows the spread of temperature (or density) experienced by the disc at that radius throughout its entire lifetime. The box plot whiskers denote the 5th-95th percentile of the data. Surface densities are calculated for the discs after they have been rotated to be face-on, so that the orientation of the disc has no effect.}
    \label{fig:tempSurf}
\end{figure*}

Figure \ref{fig:tempSurf} depicts these distributions, where each box-whisker shows the aforementioned ranges of temperature and surface density. The temperature profiles of the $\gamma_1$ discs follow a similar pattern to one another; they have high temperatures at low radii that decrease rapidly, levelling out at around $10 \, \rm K$ for the remainder of the disc. On the other hand, the $\gamma_{1000}$ discs decrease in temperature much less rapidly, and instead plateau at around $\rm 100 \, K$. Increasing $\gamma_{\rm SFR}$ has increased the temperature of the discs, and kept the temperatures high even at large radii. The surface density profiles of the discs are more comparable between the clouds. The discs are densest at the centre and decrease in density with radius, until levelling off in a similar fashion to the temperature profiles. The $\gamma_{1000}$ discs generally have higher surface-density than their $\gamma_1$ counterparts at all radii and they tend to retain a surface-density above the minimum cut off for \textsc{discfind} out to a much larger radius.

The amount of variation within the radial profiles increases with $\gamma_{\rm SFR}$, especially for the surface density profile. The range of temperatures and surface densities experienced at a given radius is generally small for the $\gamma_1$ discs, whereas for the $\gamma_{1000}$ discs it is much larger. Variation in the radial profiles implies that the state of the disc at that radius changed a lot over the disc's lifetime, rather than remaining in a steady state. This variation could be due to a number of factors, such as disruption, accretion, interactions, or fragmentation. These events are more common occurrences in the $\gamma_{1000}$ clouds, explaining the larger variability in these disc profiles.

The temperatures of the $\gamma_{1000}$ discs decreasing much less rapidly than their $\gamma_{1,2}$ counterparts is consistent with them forming from hotter gas. The strengthened ISRF can penetrate further into the envelope around the disc and heat it, and enhanced cosmic ray heating heats the gas regardless of how well shielded the disc is. These effects lead to temperatures in excess of $100 \rm \, K$ for the whole disc. In contrast, weaker cosmic rays and a largely attenuated ISRF result in the $\gamma_1$ discs cooling to the dust temperature, around $10 \rm\,K$. The surface densities of the $\gamma_{1000}$ discs increase slightly as they form in denser, more massive cores. The increased temperature and increased surface density of the $\gamma_{1000}$ discs effectively cancel out when determining the Toomre parameter, allowing the discs to remain globally stable despite massive changes to their radial profiles. 

The flat temperature profile of the $\gamma_{1000}$ discs implies that external heating is the dominant mechanism for setting the discs' temperature. This may prevent the discs from being able to self-regulate, as they can no longer heat up and cool down through the formation and dissipation of gravitational instabilities. This leads to the formation of weaker and more short-lived spiral arms, as has been noted in \citet{rowther_short-lived_2024}, and seems evident in Figure \ref{fig:firstDiscsPanel}. This may in-part explain why these discs are more severely disrupted by their fragmentation events than their $\gamma_1$ counterparts, which can more easily self-regulate to avoid catastrophic fragmentation.

\section{Events in the Lifetime of a Disc} \label{sec:discEvents}

\begin{figure*}
    \centering
    \includegraphics[width=0.72\textwidth]{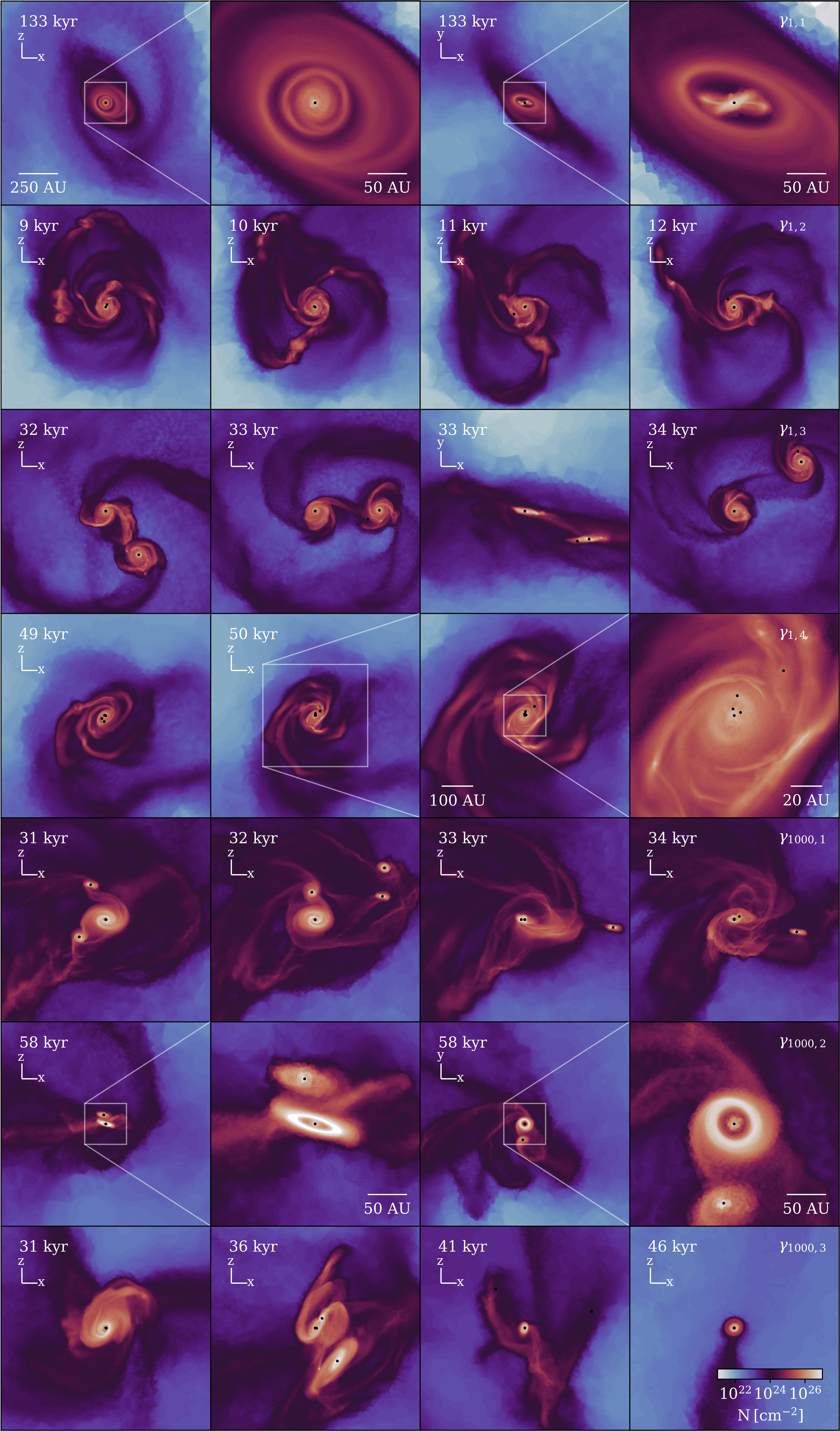}
    \caption{Column density maps of each disc at a notable point in its evolution. Each row of panels shows either a temporal sequence, a sequence of different projections at the same time, or a combination of both. All panels show the discs projected in the x-z plane of the simulation box unless otherwise stated. Spatial scales are the same as the top-left panel unless otherwise stated, and the colourmap scale is the same across all panels. Sink particle locations are shown as black dots.}
    \label{fig:interestingDiscEvents}
\end{figure*}

We showed in Section \ref{sec:discEvolution} that the individual environment around each disc leads to them each experiencing a unique evolution across their lifetimes. This leads to a wealth of diversity of events and evolutionary paths in even a small sample of 7 discs, and provides an opportunity to explore the different events that can occur in a disc's lifetime. We have collated an interesting event that occurs in each of the discs' evolution, and presented them in Figure \ref{fig:interestingDiscEvents}.

\subsection{A Misaligned Inner and Outer Disc} \label{sec:misalignment}

While the disc in $\gamma_{1,1}$ is generally quiescent, it does present an  example of misalignments between inner and outer discs. The inner structure of the disc illustrated in the first row of Figure \ref{fig:interestingDiscEvents} indicates the presence such a misalignment. The inner disc itself also presents evidence of a secondary misalignment. These kinds of misalignments are common in both simulations and observations \citep{jappsen_protostellar_2004, tsukamoto_formation_2013-1,bate_diversity_2018, ansdell_are_2020}, and occur as a result of a disc accreting material with a different angular momentum vector to what is already in the disc. This material either forms a second-generation disc that persists for many kyr, or replenishes an evolved disc with new (misaligned) material \citep{kuffmeier_misaligned_2021, pelkonen_origin_2025}. 

\begin{figure}
    \centering
    \includegraphics[width=\columnwidth]{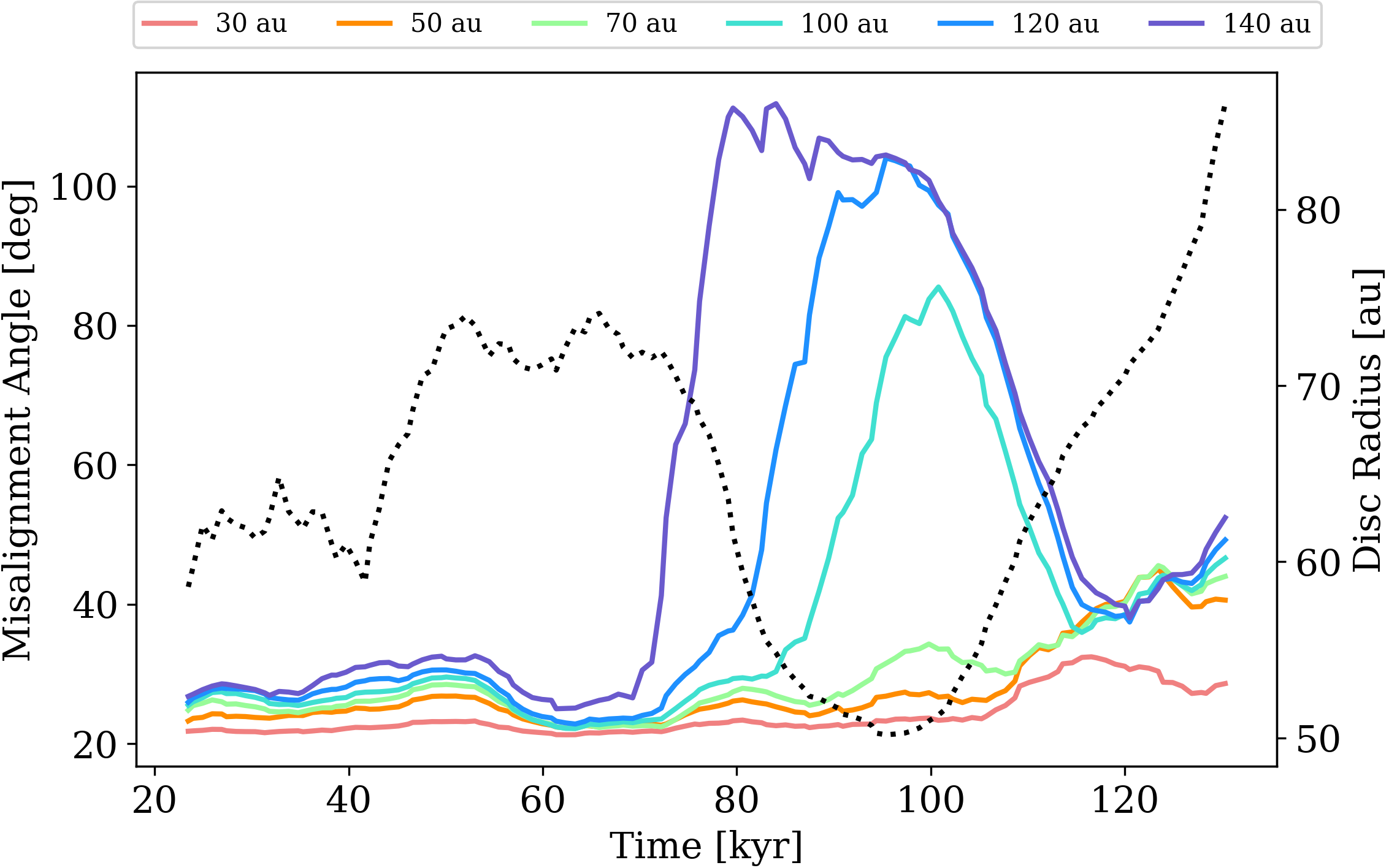}
    \caption{The angle between the angular momentum vectors of the inner and outer disc, for different outer disc radii, across time for $\gamma_{1,1}$. The inner disc is defined at a radius of 10 au. The angular momentum vector of each outer disc radius is determined by calculating the mass-weighed average angular momentum of the cells within an annulus 10 au wide around that radius. The dotted black line denotes the radius enclosing 75\% of the disc's mass (labelled on the right y-axis). Both the misalignment angle and disc radii are smoothed using a rolling average with a window of 10 kyr.}
    \label{fig:misalignmentAngle}
\end{figure}

Figure \ref{fig:misalignmentAngle} illustrates how the misalignment between the inner and outer disc varies across time, for different outer disc radii. Sometimes the disc is smaller than some of the outer disc radii we plot, and in these cases the Figure is showing the misalignment between the inner disc and the material that surrounds the disc. At early times the disc and its environment are well aligned, as the disc has formed from its surrounding material. 

After around 50 kyr, misalignment at the outermost radii rapidly increases while the disc's radius shrinks, indicating that new material with a strongly misaligned angular momentum vector has entered this disc's local region and truncated the disc. After a few kyr, the disc begins to accrete from this material and increases in radius again. As this occurs, the outermost annuli become better aligned with the disc as the new misaligned material changes the alignment of the disc itself. This new material does not entirely penetrate the inner disc, however, which remains slightly misaligned with respect to the rest of the disc by the end of the simulation. The alignment of the material with radius $R$ > 50 au is well defined, but different from the alignment of the material in the inner disc. 

This example shows how the accretion of material can drastically change the form and structure of a disc. Even a stable disc with modest accretion rates can undergo this process. The subsequent misalignment can also persist for many kyr. We follow the disc for about $\rm 30 \, kyr$ after the misalignment forms, during which time it persists. 

Accretion of the material onto the disc does weaken the misalignment. The gas causing the change begins with a misalignment of nearly 90\textdegree $\,$to the inner disc, yet produces a lasting misalignment of only 20\textdegree $\,$after it has been accreted. This suggests that during the accretion of this material onto the disc the misalignment has been reduced, either by the material losing angular momentum, or by some of the misaligned material making it into the inner disc and changing the angular momentum vector there as well. This highlights the complexity of these misalignments.

\subsection{A Chaotic Unstable Disc} \label{sec:chaoticDisc}

Figure \ref{fig:discViscosityEvolution} demonstrates that, formally, the discs are Toomre-stable for their entire lifetimes. While overall the discs may be stable, locally this may not be the case. This can be quantified by instead calculating the Toomre parameter locally, for all regions of the disc. The disc in $\gamma_{1,2}$ is a good testbed for this, as it shows evidence of strong GI, it fragments multiple times, and is relatively isolated. The second row of Figure \ref{fig:interestingDiscEvents} depicts the disc at four consecutive times separated by 1 kyr, where numerous clumps are visible in and around transient spiral arms. The inner region of the disc is typical, dominated by high-$m$ modes, while the outer regions are highly variable and made up of extended spiral arms broken up by numerous clumps. The disc also fragments after 11 kyr, at the edge of the regular portion of the disc. 

\begin{figure*}
    \centering
    \includegraphics[width=\textwidth]{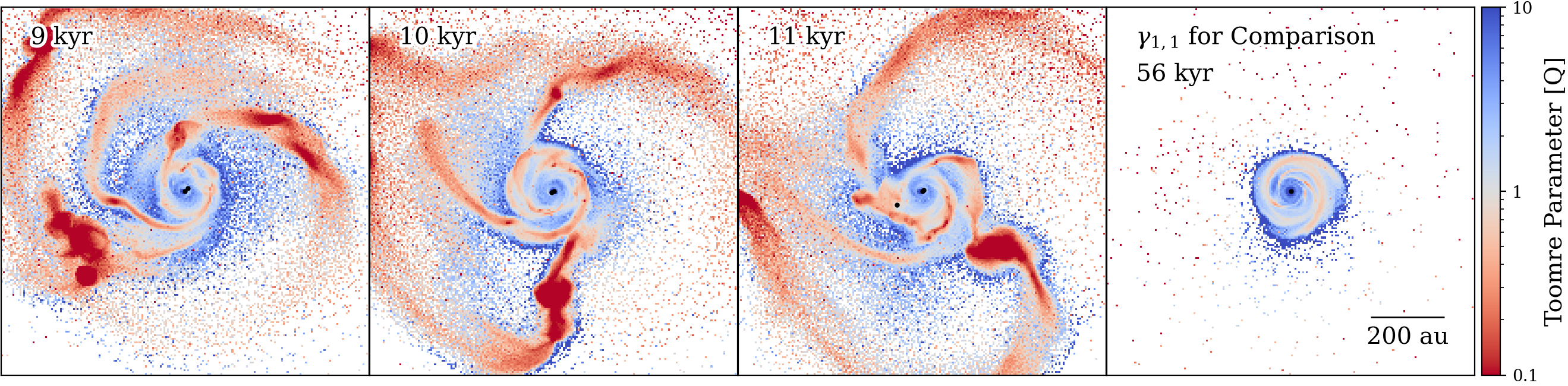}
    \caption{The Toomre stability parameter, $Q$, for the $\gamma_{1,2}$ disc, at three points in time. The region containing the disc is subdivided into bins of size 8 $\rm au^2$, for which a mass-weighted average value of $Q$ is determined. The disc shows the largest $Q$ values in the outer spiral arms, and lowest in the inner disc. The rightmost panel shows the $\gamma_{1,1}$ disc for comparison. The locations of sink particles are marked by black dots. All panels are shown at the same spatial scale.}
    \label{fig:disc2stability}
\end{figure*}

Given that the disc experiences GI and fragments, it is expected that a Toomre stability map of the disc should exhibit corresponding areas of low-$Q$. In Figure \ref{fig:disc2stability} we present such a map of the disc after 9, 10 and 11 kyr, along with a comparison map of the $\gamma_{1,1}$ disc. Instead of appearing globally stable the disc now contains many regions of local Toomre-instability, such as the spiral arms or clumps that have formed at the edges of the disc, though much of the inner and inter-arm regions of the disc still appear stable. 

The disc fragments between 10 and 11 kyr, just outside its stable inner region. It does not form from the highly Toomre-unstable clumps in the outer disc, which persist across all three panels without fragmenting. It does form from gas that is marginally Toomre-unstable in the 10 kyr panel. Therefore fragments will form in gas that is unstable, but not all unstable gas will fragment, in agreement with previous studies of isolated or two-dimensional discs.

The disc of $\gamma_{1,1}$ instead presents a stark contrast, where almost no regions of the disc display a low value of $Q$. While the globally-averaged Toomre parameters of the discs are similar, locally the maps of $Q$ look completely different. A disc with a large average value of $Q$ can have many regions that are locally Toomre-unstable, and which may even fragment under the right conditions. This supports the notion that this parameter is not sufficient on its own to assess whether the whole disc is stable, or whether it will fragment. It also indicates the complexity of fragmentation in discs, as any condensing proto-fragment needs to overcome both kinetic and rotational pressure. Our current analysis assumes an equilibrium disc, i.e one where there is no turbulent velocity dispersion in addition to thermal support and the disc's rotation curve is perfectly Keplerian, for which the Toomre parameter is valid. When the discs leave equilibrium, as they do in this example, they enter a regime where the Toomre criterion is no longer a valid condition for fragmentation on its own. 

\subsection{Flyby-Induced Fragmentation} \label{sec:discInteraction1}

In a star-forming environment it is rare that a core spawns only a single star, and instead stars are born in small groups \citep{goodwin_limits_2005,bate_stellar_2009,lomax_simulating_2014, ambrose_formation_2024}. Each of the stars in these groups will have its own accretion disc, and these can interact with one another. This occurs in the case of $\gamma_{1,3}$ (Figure \ref{fig:interestingDiscEvents}, third row), where it experiences a close flyby with another disc, which appears to induce a fragmentation event.

The two sink particles pass within $350\rm \, au$ of each other at their closest approach (Figure \ref{fig:interestingDiscEvents}, third row, rightmost panel). This is within, but on the higher end of, the 100 to 400 au encounter distance condition put forward by \citet{cadman_binary_2022} for interactions to trigger fragmentation. The edges of the discs interact with each other, the outermost spiral of the left-hand disc extending out and down to the other disc. By this point, however, the disc has already fragmented. The projection of the system in the third panel of Figure \ref{fig:interestingDiscEvents} shows that the discs do not interact as closely as they appear to in a top-down view. Given the clumps visible in the passing by disc in the leftmost panel of the Figure, it is possible that this disc was going to fragment whether a flyby occurred or not. 

\subsection{A Complex Disc Central Cluster} \label{sec:discCluster}

The fragmentation of the disc in $\gamma_{1,4}$ presents a case study of the evolution of low-mass fragments that form in a typical disc. After $\rm 50 \, kyr$ of evolution the disc has fragmented 5 times, and each fragment has migrated to the inner regions of the disc, as shown with successive zooms in the fourth row of Figure \ref{fig:interestingDiscEvents}. 

\begin{figure}
    \centering
    \includegraphics[width=\columnwidth]{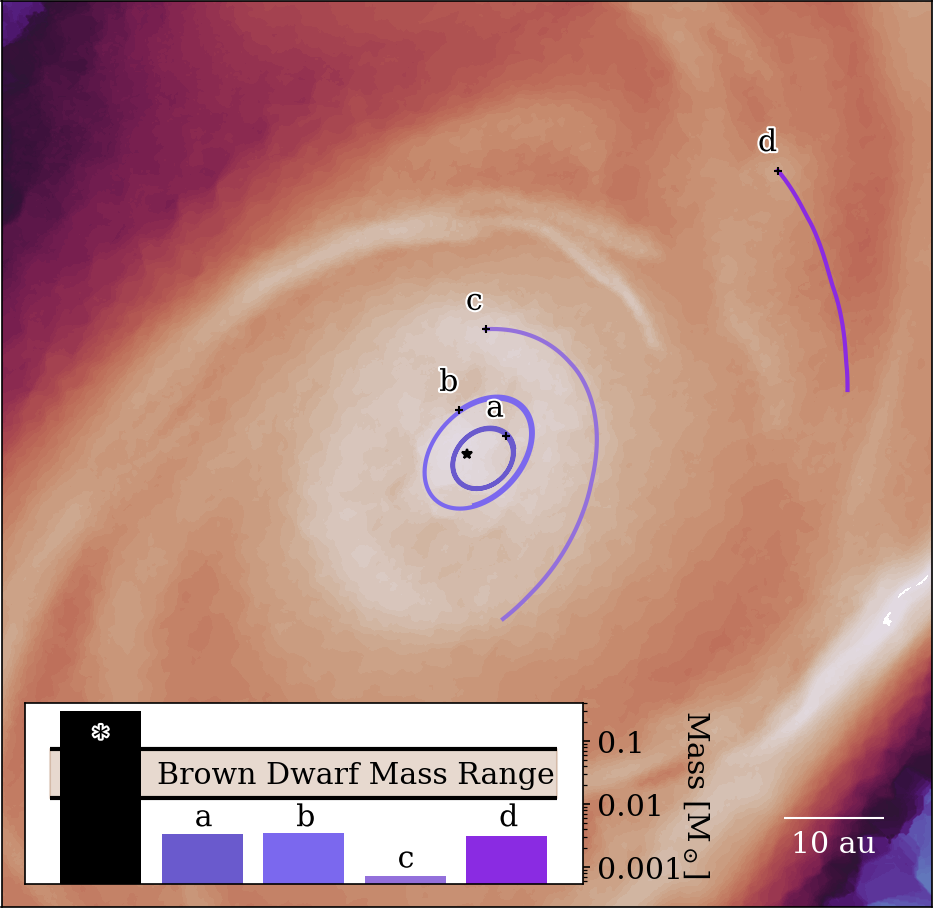}
    \caption{The positions and trajectories of the sinks in the central region of $\gamma_{1,4}$'s disc. The trajectories are drawn from a 50-year purely N-body integration of the sink particles. The first sink is plotted as a star symbol with no accompanying label and the companion sinks are plotted as plus symbols. The inset axes show the masses of each of the sink particles.}
    \label{fig:discCluster}
\end{figure}

Figure \ref{fig:discCluster} shows the positions and expected trajectories of each of the sink particles in the cluster, along with the masses of each of the sinks. The central object, denoted "*", is a low-mass star and dominates the system. The companions are all planetary-mass objects, ranging from 1 to 4 $\rm M_{Jup}$, and appear to have stable orbits around the central sink. These masses are on the low end of what is expected from disc fragmentation, but still consistent with the result of various studies that giant planets can form from disc fragmentation \citep{rice_substellar_2003, stamatellos_properties_2009, kratter_runts_2010, forgan_jeans_2011, forgan_towards_2013, hall_identifying_2017}. Therefore these fragments could persist and evolve into a fast-orbiting giant planet such as a hot Jupiter once the disc has dissipated. 

Since the fragments did not form in their current location (see Section \ref{sec:discFragmentation}), they must have migrated inwards after formation. It is plausible that they could continue to do so, as the stable orbits seen in the 50 year N-Body simulation in Figure \ref{fig:discCluster} do not take into account the dissipative effect of the disc's gas on the sinks. Therefore the fragments could migrate further inwards and merge with the central star, likely producing an FU-Orionis type event, as has been seen in numerous simulations \citep{vorobyov_burst_2006,vorobyov_formation_2010,machida_recurrent_2011, vorobyov_formation_2013,tsukamoto_formation_2013,vorobyov_gravitational_2018}. However, since we do not model merging of sinks in these simulations, we cannot say which scenario would occur. We can confirm, however, that such objects can form from disc fragmentation and that they tend to migrate to the inner regions of the disc after forming. 

\subsection{Flyby-Induced Disruption} \label{sec:discInteraction2}

The event shown for disc $\gamma_{1000,1}$ is an example of how easily a disc can become disrupted by a flyby event. The first sink and its disc are fairly massive, at around $\rm 1.75 \, M_\odot$ and $\rm 0.9 \, M_\odot$ respectively. The sink has a companion (to its lower-left of the first panel of the fifth row of Figure \ref{fig:interestingDiscEvents}), with its own disc. Both of these are light, with masses $\rm 0.28 \, M_\odot$ and $\rm 0.08 \, M_\odot$ respectively. The system is approached by a third sink (to the upper left of the first panel of the fifth row of Figure \ref{fig:interestingDiscEvents}), with a disc and sink mass roughly half that of the initial companion. Despite the difference in mass between the initial and perturbing system, the most massive disc is completely disrupted by the interaction. The primary and secondary sinks in-spiral, forming a tighter binary system with the remains of the two initial discs forming disordered circumbinary material. The system remains highly irregular for a long time ($\rm > 30 \, kyr$) and never settles back down to the type of disc seen before the interaction. 

This example shows how a perturbing disc or sink need not have a mass of the same order of the disc it disrupts. A relatively light interloper is capable of destroying a comparatively massive disc, and a larger disc mass does not protect a disc from disruption. In a dense, clustered, environment like the $\gamma_{1000}$ clouds this means that discs are unlikely to remain undisturbed or regular for very long. 

\subsection{A Ring Disc}

The disc in $\gamma_{1000,2}$ shows a ring-like morphology, where instead of a contiguous circumstellar disc there is both an evacuated cavity around the central sink and a sharp cut-off point at the edge of the disc. In this instance, the cavity is approximately $\rm 23 \, au$ in radius and the ring itself is around $\rm 13 \, au$ wide. This type of structure is different to protoplanetary rings seen in observations, where cavities are formed by accreting planetesimals evacuating their local area of material \citep{andrews_protoplanetary_2021, carrera_protoplanetary_2021}. Instead, this ring is a result of disc fragmentation and the resulting shape of the local potential.

\begin{figure}
    \centering
    \includegraphics[width=\columnwidth]{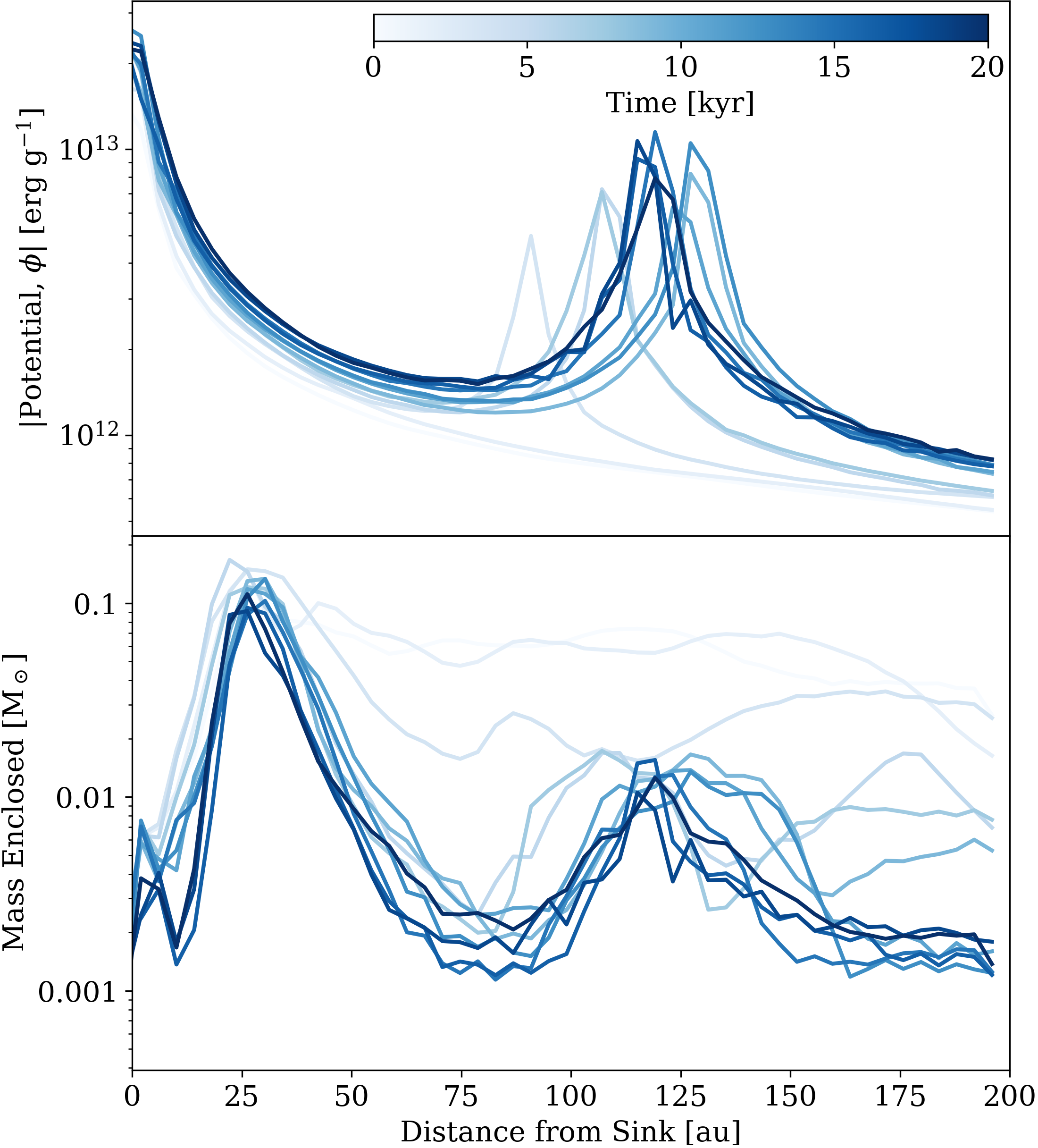}
    \caption{The evolution of the shape of the gravitational potential and mass around the $\gamma_{1000,2}$ sink shortly before and after the fragmentation event. The data are binned into annuli 2 au wide, within which the maximum potential and mass enclosed are determined from the cells within the annulus. The apparent discrepancy between the top and bottom panels at low radii is due to the central sink not appearing in the mass distribution. The sink's mass induces a potential peak while the mass distribution falls to a minima. The combination of both the gas and sink mass produces the potential seen.}
    \label{fig:ringPlot}
\end{figure}

Before the disc in $\gamma_{1000,2}$ fragments, the potential map in Figure \ref{fig:ringPlot} is centrally peaked as a result of the distribution of the disc mass and presence of the central sink. The radial mass distribution, shown in the bottom panel of the same Figure, behaves similarly. Once the disc fragments a new peak in the potential appears at a radius of around 125 au, the location of the fragment. As the fragment grows and accumulates its own disc, the potential well deepens to become comparable to the central sink. This produces a "u" shape in the potential. Any gas located at the centre of that "u" is in an unstable position, as any perturbation that causes it to lose angular momentum will lead it to "fall down" into either one of the two potential wells. Over time this evacuates the cavity between the two sinks, as seen in the mass distribution evolution. The widths and depths of the potential wells created by the two sinks sets an outer limit to the size of the discs around the sinks, explaining the outer radius of the ring disc. 

The inner region of the ring disc also contains little gas, but not as a result of the potential. In most of the discs in this set of simulations there is a central cavity. These are the result of the central sink "wobbling" around the common barycentre of the star-disc system. This wobble can be very pronounced in the $\gamma_{1000}$ discs, as they are massive and of comparable mass to their host star. The sink clears the gas within this orbit, leaving behind a central cavity larger than the size of the sink itself. After fragmentation and disruption of the disc this cavity remains, producing the thin ring-shape. 

\subsection{Flyby-Induced Ejection}

Similar to $\gamma_{1000,1}$, the disc of $\gamma_{1000,3}$ undergoes a flyby event. However, instead of disruption of the disc this flyby results in the star being ejected from its natal core very early, taking its remaining disc with it. The system is initially made up by a large disc on the order of $\rm 1 \, M_\odot$, which is slightly larger than its host binary system that recently formed through the disc fragmenting. Shortly after $\rm 31 \, kyr$ (Figure \ref{fig:interestingDiscEvents}, first panel, bottom row) the disc fragments again, forming a triple system. Following this, a secondary star-disc system flies in from the bottom of the panel, passing within $\rm 300 \, au$ of the disc. This quadruple-body interaction strongly disrupts the system, ejecting all three of the original disc's sinks in different directions and leaving the interloper disc undisturbed.

This kind of "sling-shot" interaction is common in N-body simulations, where such events are responsible for the presence of lone stars \citep{ambrose_formation_2024, polak_massive_2024}. This event confirms that such events still occur in fully-hydrodynamic models where circumstellar discs have formed. It also highlights how ejected stars can retain their discs, despite leaving their natal core and undergoing a disruptive interaction. The disc of the ejected star is much lighter than it was before the interaction, now $\rm < 0.1 \, M_\odot$, and shows evidence of a comet-like tail, but it is still present. Eventually the sink accretes the disc fully and it becomes undetectable to \textsc{discfind}, as it can no longer maintain its mass through replenishment from the surrounding core.

\section{Disc Fragmentation} \label{sec:discFragmentation}

Some of the events shown in Section \ref{sec:discEvents} are related to, or occur as a result of, disc fragmentation. The process of disc fragmentation is a crucial part of the evolution of circumstellar discs, and much work has been done to understand its prevalence, role and products \citep{stamatellos_properties_2009, vorobyov_fragmenting_2013,kratter_gravitational_2016}. In particular, disc fragmentation is likely to help populate the low mass end of the initial mass function by forming brown dwarf and planetary mass objects. Therefore it is pertinent to investigate how disc fragmentation may differ in different environments. 

\subsection{Fragmentation Maps}

 We have previously shown both local and global Toomre maps, and found them to be insufficient to predict fragmentation on their own. However, it is generally accepted that Toomre instability is not the only condition that needs to be met for fragmentation to occur. The use of $Q$ as a condition for fragmentation is also complicated by the effects of infall. Accretion onto the disc raises the local surface density, which can artificially lower $Q$. This can occur without the disc experiencing the GI growth or increased angular momentum transport associated with a low $Q$ \citep{tsukamoto_effects_2015}.

 In addition we should also consider that the parcel of gas that is going to fragment needs to be able to cool and collapse on a timescale shorter than the dynamical timescale of the disc, lest it be torn apart by shearing forces. This can be parametrised by the $\beta$ parameter \citep{gammie_nonlinear_2001},

\begin{equation}
    \beta = \tau_{\rm cool}\,\Omega(R) \, ,
\end{equation}

\noindent where $\tau_{\rm cool}$ is the local cooling time and again $\Omega(R)$ is the angular speed as a function of radius. Parcels of gas with a low value of $\beta$ have a cooling timescale shorter than the dynamical timescale, allowing them to collapse. The exact value of $\beta$ needed for fragmentation to occur is unclear, and appears to depend on numerous factors such as the adiabatic index and cooling prescription \citep{rice_investigating_2005, cossins_effects_2010}, irradiation \citep{rice_stability_2011, kratter_fragment_2011}, disc surface density profile \citep{meru_fragmentation_2011}, numerical setup \citep{meru_non-convergence_2011} and the thermal history of the disc \citep{clarke_response_2007}. Therefore it may be more instructive to compare the $\beta$ value relative to other locations in the same disc, rather than to a specific critical value.

\begin{figure*}
    \centering
    \includegraphics[width=0.90\textwidth]{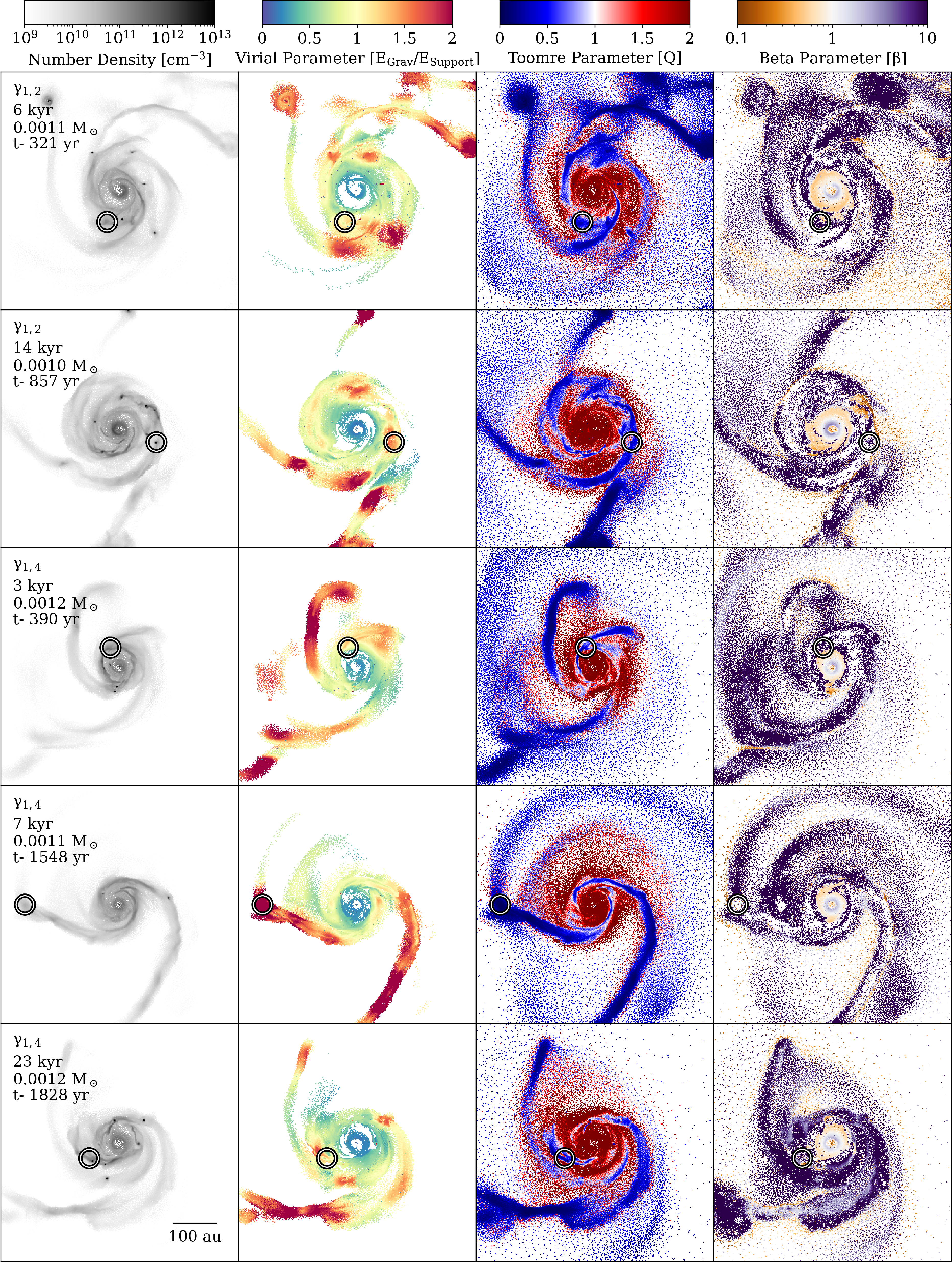}
    \caption{Maps of the number density, local virial parameter, Toomre $Q$ parameter and beta parameter of the $\gamma_1$ discs at times just before they fragment. The region containing the cells that will form the fragment is circled. The disc that fragments, time of the fragmentation, the initial mass of the fragment and the time differential between the map's creation and sink formation is shown in the top-left of each number density panel. The maps are generated by first discretising the disc into 1 au x 1 au "pixels" in the plane of the disc, which extend up to 50 au above and below the disc. Within these pixels mass-weighted quantities are determined using the cells present within that pixel. The procedure for the calculation of the local virial, Toomre and $\beta$ parameters is described in Appendix \ref{sec:mapquants}.}
    \label{fig:fragmentationMaps1}
\end{figure*}

\begin{figure*}
    \centering
    \includegraphics[width=0.90\textwidth]{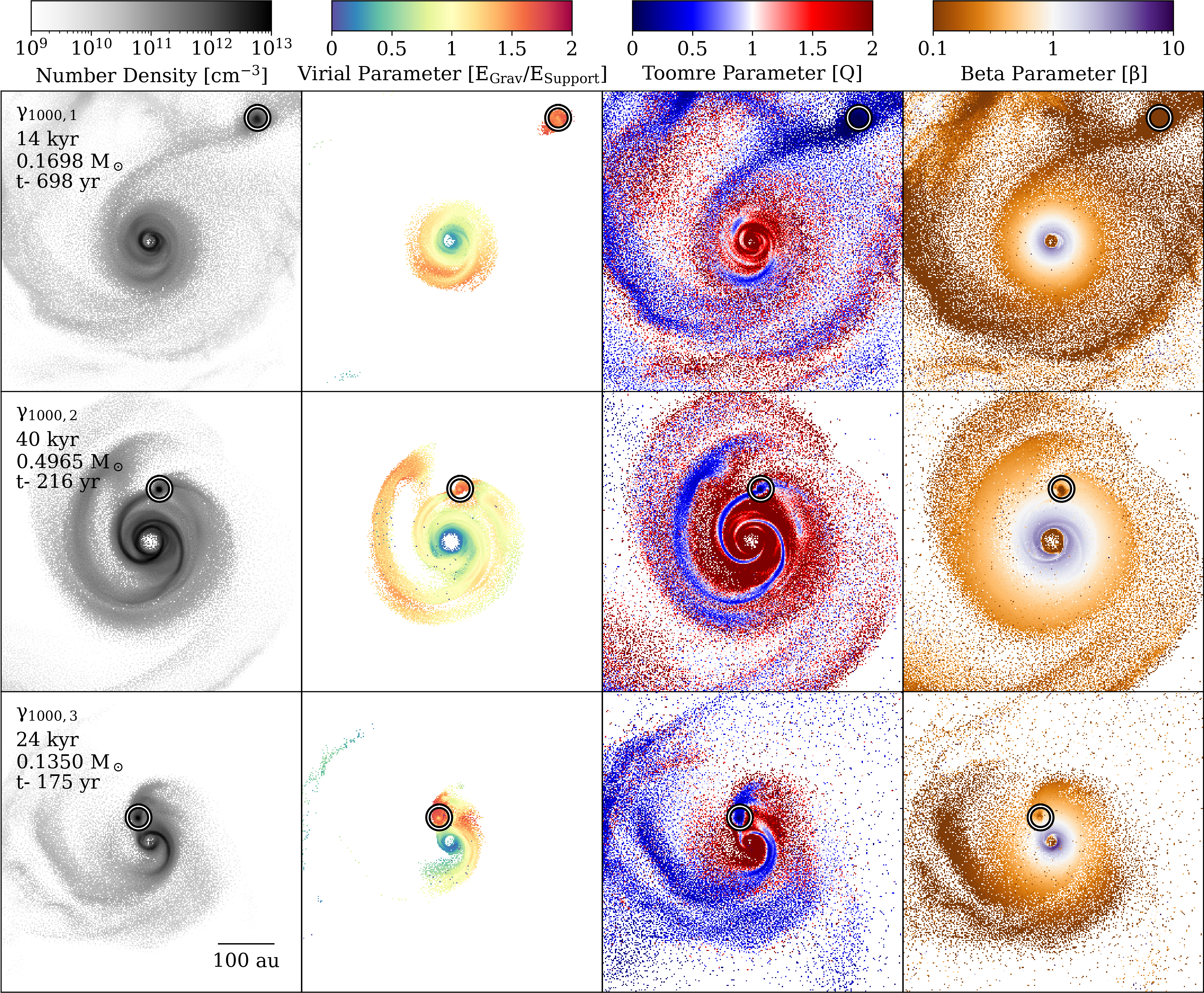}
    \caption{As Figure \ref{fig:fragmentationMaps1}, but for fragmentating discs in the $\gamma_{1000}$ simulations.}
    \label{fig:fragmentationMaps2}
\end{figure*}

Figures \ref{fig:fragmentationMaps1} and \ref{fig:fragmentationMaps2} depict maps of 
number density, local virial parameter, Toomre stability and $\beta$ for a representative sample of the fragmentation events that occur in the first discs\footnote{Appendix \ref{sec:mapquants} details how the quantities in the maps are determined.}. Figure \ref{fig:fragmentationMaps1} indicates the events that occur in the $\gamma_1$ discs, and Figure \ref{fig:fragmentationMaps2} the events that occur in the $\gamma_{1000}$ discs.

\subsubsection{The $\gamma_1$ Events}

Of the four $\gamma_1$ discs, two of them fragment. The discs that do fragment do so multiple times, typically in succession with around 5 to 10 kyr between events. This may be an example of triggered fragmentation \citep{meru_triggered_2015}, where changes to the dynamical behaviour of the discs due to fragmentation facilitates further fragmentation. However the exact mechanism proposed by \citet{meru_triggered_2015} suggests that subsequent fragmentation will occur at smaller radii, which is not generally the case here. The fragments instead tend to form in the outer regions of the discs even after previous fragmentation events. They do typically migrate to the centre of the disc, as seen in Section \ref{sec:discCluster}, but they do not form there. 

The maps of number density show the presence of numerous dense clumps throughout the discs. These clumps are generally spread across the mid-to-outer regions of the discs. While dense, these clumps are not always gravitationally bound and do not necessarily show up in the virial parameter map. These clumps are also rarely the progenitor of fragments. Despite being notable features of the discs, they are instead transient structures that seldom persist for more than a few kyr. These clumps are common in disc simulations \citep{hall_identifying_2017} and are the result of interacting high-$m$ spiral modes \citep{xu_globall_2025}. 

The regions of the disc that do fragment are not easily discernible from the maps in Figure \ref{fig:fragmentationMaps1}. Only one event, $\gamma_{1,2}$, is associated with a dense, bound, clump. The other regions are generally dense and marginally bound, but not anomalously so compared with their surroundings. Up to 2 kyr elapses between the calculation of these maps and the formation of the sink particles, which is multiples of both the orbital and free fall timescales of the fragmenting regions. At times closer to the formation of the sink the regions would likely become more distinguishable in the maps, such as that of $\gamma_{1,2}$.

The regions of the disc that appear the most locally bound are the outermost spiral arms, or denser sections of the inner arms. The inner disc is generally locally unbound, as the shearing forces are strongest there. This is where the Toomre parameter is largest too, for similar reasons. The disc is Toomre-unstable in its spiral arms and the outer portions of the disc. The fragmenting region is always Toomre-unstable, but it is not unique in being so (as also seen in Section \ref{sec:chaoticDisc}). Finally, the $\beta$ parameter is large for almost the entire disc, and is large for each of the fragmenting regions. This suggests that the bulk of the disc should not be able to cool rapidly enough to overcome shearing forces and collapse. However the $\beta$ parameter is calculated from the current cooling rate of the gas; as the fragmenting regions continue along collapse, it is likely that the $\beta$ parameter would adjust in response to the thermodynamic state of the gas changing during contraction. 

There appears to be an anti-correlation between the Toomre and $\beta$ parameters in the central region of the discs: they are highly Toomre-stable, yet the cooling times are very short. The inner disc is hot, dense, and rapidly rotating. This drives up the sound speed and epicyclic frequency, increasing the Toomre parameter. Conversely the high temperatures drive down the cooling time, reducing $\beta$ in turn. Therefore, despite being able to cool sufficiently to fragment the shear in the inner disc is too high for collapse to occur.  

The conditions for fragmentation, quantified by the virial, Toomre, and $\beta$ parameters are not met as much as 2 kyr before fragmentation in these discs. This indicates that fragmentation must occur quickly, as a result of rapid changes to the state of the disc. It also suggests that these parameters cannot be used to predict where and when fragmentation will happen, at least not within a few hundred years of it occurring. This is, in part, a consequence of our temporal resolution far exceeding the free-fall time of the gas at the sink creation density (< 0.01 kyr). Sink particles can form very quickly, and in a dynamically evolving disc the predictors of this formation may not appear on timescales greater than our temporal resolution, leading to them being missed. This suggests that using such parameters to predict disc fragmentation will always be difficult, as one would need to analyse the disc so close to the fragmentation event that its location and the question of whether or not fragmentation is going to occur would already be answered without any additional analysis.

\subsubsection{The $\gamma_{1000}$ Events}

Each of the $\gamma_{1000}$ discs fragments exactly once, a few tens of kyr into its lifetime. The fragments that form are massive, becoming stars in their own right and acquiring their own discs. As a result of this the original disc is strongly perturbed.

Instead of small, transient clumps, the $\gamma_{1000}$ discs show one significant overdensity each: the forming fragment. These discs are dominated by low-$m$, rather than high-$m$ spiral modes, rendering the mechanism that formed low mass, transient, clumps in the $\gamma_1$ discs inactive. The increased Jeans mass within these discs (as a result of their higher temperatures and densities) also forces accumulations of mass to grow much more massive before becoming bound, further preventing the formation of low mass clumps. Similarly the local Toomre mass will likely be larger in these discs as it scales rapidly with the local velocity dispersion, which is increased in the $\gamma_{1000}$ case due to the discs' elevated temperatures.

The clumps that do form stand out strongly in the maps of the virial and beta parameters, and are all Toomre-unstable. They are all on the cusp of being gravitationally bound, and show very low $\beta$ parameters, indicating that these fragments are on the way towards collapse. Unlike the fragmenting regions in the $\gamma_1$ discs, those in $\gamma_{1000}$ are visible in the maps up to $700$ years before the sinks form. The $\gamma_{1000}$ clumps need to accumulate more mass before collapsing, due to the larger Jeans mass in the disc, which increases the time that the clump will be visible in the disc. Whereas fragmentation appears to be a rapid process in the $\gamma_1$ discs, it is not so for the $\gamma_{1000}$ discs. This may make signatures of fragmentation visible for longer in such environments. 

Most regions of the discs are not bound to themselves; the only areas with a high virial parameter are the fragments. Unlike the $\gamma_{1}$ discs, these discs' spiral arms are not bound to themselves. The Toomre-parameter again shows stability in the inner disc and inter-arm regions, and instability in the spiral arms. Less of the discs overall appear to be Toomre-unstable, compared to those in $\gamma_1$. The $\beta$ parameter is low for the majority of the discs, only becoming large at the very centre. This indicates that large portions of the discs could cool quickly enough to collapse, yet do not do so. These portions, however, are not gravitationally bound like the fragment and thus cannot collapse.

\subsubsection{A Different Mode of Fragmentation}

The difference in fragmentation between the $\gamma_1$ and $\gamma_{1000}$ discs can be traced to their different properties. \citet{forgan_effect_2013} found a dependence of fragment size on surface density, and that irradiated discs produce larger fragments. Thus, given that the $\gamma_{1000}$ discs are both hotter and denser (see Figure \ref{fig:tempSurf}), it is expected that they would form more massive fragments. The increase to temperature and density in these discs increases the Jeans mass, that is the mass fragments need in order to collapse. The $\gamma_1$ discs, in contrast, have low Jeans masses and clumps can collapse much more easily. The interaction between high-$m$ spiral modes in these discs can easily accumulate this mass, driving the formation of low-mass clumps. In the $\gamma_{1000}$ discs low-$m$ modes dominate, reducing this channel of clump formation, and the large Jeans mass reduces the likelihood of such clumps persisting. Instead, these discs can only form the massive fragments we see.

The change in temperature and density of the $\gamma_{1000}$ discs also affects the $\beta$ parameter. As the discs temperatures increase, the corresponding cooling time becomes shorter due to the strong scaling of dust cooling with temperature. All the discs are in a density regime (see Appendix \ref{sec:discDensities}) where gas and dust temperatures and coupled, such that dust temperatures in the $\gamma_{1000}$ are also significantly elevated. Hotter dust radiates away the same increase in energy much more rapidly than cooler dust, and since gas-grain interactions are the dominant cooling mechanism in high-density gas this shortens the cooling time of the hotter discs. \citet{kratter_fragment_2011} argue that this can lead to enhanced fragmentation, as the $\beta$ parameter will always be low in such discs. This is also true for our discs, which have low $\beta$ values across most of their radii. The $\gamma_{1000}$ discs fragment only once each, and do so after a longer period of time, so it is difficult to say whether they fragment more easily. There is a balancing act between the change to the cooling time, allowing fragments to collapse, and the change in Jeans mass, requiring fragments to grow more massive before collapsing. The result of this is that large, singular fragments form in the $\gamma_{1000}$ discs. On the other hand, the large values of $\beta$ across the $\gamma_1$ discs may be preventing larger fragments from forming, as they are sheared apart before they can cool and collapse, leaving behind only a light core. The lighter Jeans mass in these discs also means that the clumps that do form are naturally less massive as well. 

\subsection{Disc Accretion Rates}

Accretion is a key environmental factor that can play a pivotal role in disc fragmentation by triggering gravitational instability \citep{kuffmeier_zoom-simulations_2017, kuffmeier_episodic_2018}. Discs with infall are more likely to experience global angular momentum transport as the viscosity of the disc increases to process the new material \citep{vorobyov_self-regulated_2007, harsono_global_2011}. Discs that accrete strongly will necessarily grow larger and denser, and variation in the accretion rate can push the disc to gravitational instability as self-regulation breaks down. \citet{kratter_role_2010} and \citet{offner_formation_2010} suggest that in the context of disc fragmentation (also see \citet{shu_self-similar_1977} and \citet{girichidis_importance_2011} for the context of star formation more generally) the parameter 

\begin{equation}
    \xi = \frac{G\dot{M}}{c_s^3}
    \label{eq:maxAcc}
\end{equation}

\noindent cannot exceed 2 to 3 without a disc fragmenting, where $\dot{M}$ is the infall rate and $c_s$ is the sound speed in the disc. 

\begin{figure}
    \centering
    \includegraphics[width=\columnwidth]{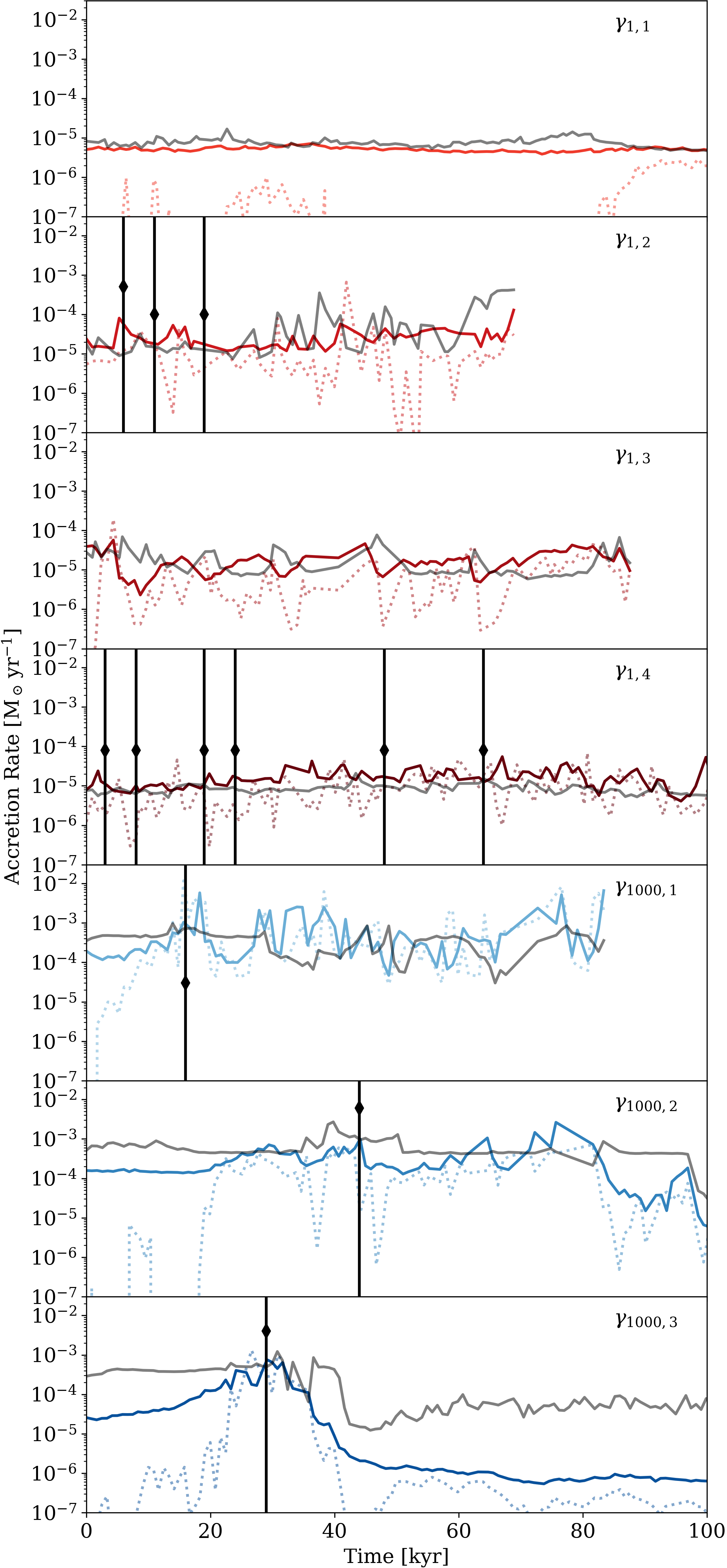}
    \caption{Accretion rates onto each of the discs against time. Accretion rates are calculated by taking the sum of the mass flux at a radius of $\rm 250 \, au$ from the central sink. The sum of the positive (onto the disc) accretion is shown as a solid line, and the sum of the negative (away from the disc) accretion is dotted. Times when the discs fragment are denoted with vertical lines and diamonds. The solid grey line represents the theoretical maximum accretion rate that the discs can process without fragmenting, from \citet{offner_formation_2010} and \citet{kratter_role_2010}, calculated using Equation \ref{eq:maxAcc} assuming $\xi = 2.5$.}
    \label{fig:accretionTimelines}
\end{figure}

Figure \ref{fig:accretionTimelines} portrays the accretion rate onto the discs across their lifetimes. Most of the discs experience a large amount of variability in their accretion rate, which sometimes spans up to 3 orders of magnitude. The $\gamma_{1000}$ discs, like their host stars, experience higher accretion rates, averaging around $10^{-4} \rm \, M_\odot \, yr^{-1}$ compared to $10^{-5} \rm \, M_\odot \, yr^{-1}$ for the $\gamma_1$ discs. These values are consistent with the accretion rates onto the sinks, indicating that the discs are generally processing the material they accrete and feeding it to the host star at a stable rate rather than growing significantly. These accretion rates are also typical of those seen in other simulations, such as those in \citet{mauxion_modeling_2024} and \citet{mayer_protostellar_2025}. Both sets of discs also undergo periods of "negative" accretion, where the net mass flow is away from the disc rather than towards it. These periods are also not generally associated with fragmentation events, and instead appear to occur largely at random. 

Both the $\gamma_1$ and $\gamma_{1000}$ discs tend to accrete at around the theoretical "limit" for most of their lifetimes. Only when the $\gamma_{1000}$ sinks are ejected does their accretion rate fall significantly below the limit. There appears to be no relationship between the accretion rate onto the disc and fragmentation. It is also unclear if accretion above the theoretical limit causes fragmentation. In $\gamma_{1,4}$, prolonged accretion above the limit coincides with recurrent fragmentation, but in $\gamma_{1,3}$ it does not. The catastrophic fragmentation events in the $\gamma_{1000}$ do not generally occur when the disc is accreting above the limit either. Accretion above the limit may instead perpetuate gravitational instability, rather than fragmentation, as all the discs experience recurrent GI across their lifetimes coincident with accretion at or above the limit.

\section{Discussion}

\subsection{Disc Properties Depend on Environment}

The properties and evolution of the discs differ greatly between the $\gamma_1$ and $\gamma_{1000}$ clouds. These differences are a consequence of how increasing $\gamma_{\rm SFR}$ affects the structure and thermodynamics of molecular clouds. Many of these changes, such as larger core masses, differing core distribution, stronger accretion onto stellar systems, were described in \citet{cusack_fragmentation_2025}. We have illustrated that those changes cascade down to the disc-scale and produce discs that are hotter, denser, more massive and rapidly accreting, compared with those in a solar-neighbourhood like environment. Similarly their host stars accrete more rapidly and grow to become more massive than their low-$\gamma_{\rm SFR}$ counterparts. 

The $\gamma_{1000}$ cloud is hotter and generally denser than that of the $\gamma_1$ cloud. This increases the Jeans mass in the cloud, leading to the formation of more massive cores. This provides a larger mass reservoir for the stars and discs that form within the cores, allowing them to grow more massive. The stronger ISRF and CRIR in these clouds also results in higher temperatures in the discs, which in turn allow them to grow denser before reaching instability. In the $\gamma_1$ clouds the comparatively weak ISRF is attenuated before reaching the discs, and the low CRIR does not provide sufficient heating to increase their temperatures. This prevents the discs from becoming too dense, as they will reach instability much more quickly when at low temperatures. 

\subsection{Disc Evolution Depends on Environment}

In addition to the thermal state of the environment setting the temperature, density and mass of the discs, many other environmental factors impact how a disc evolves. Discs are not born in isolation, and instead are joined by other discs, protostars, streamers and other structures within their natal core. The interactions between these is key to the evolution of a disc. We highlighted in Section \ref{sec:discEvents} a disc can undergo a wide range of (often impactful) events that are all the result of interactions with the disc's environment. We also demonstrated in Figure \ref{fig:firstDiscsPanel} that such discs can display a wide range of morphologies, and evolve between them over short periods of time. These morphologies are in turn influenced by their environment. 

Accretion of material onto discs occurs through streamers, affecting the structure and stability of the disc \citep{smith_quantification_2011, kuffmeier_episodic_2018, kuffmeier_misaligned_2021, calcino_anatomy_2025}. We see direct evidence of this in our discs, through the formation of misaligned components and strong accretion events that sometimes accompany disc fragmentation. Interactions between discs are common, and can trigger instability, or lead to ejection of a protostar and its disc from the natal core. Disc fragmentation can produce varied structures in the discs, such as thin rings around binary stars, or disrupt them completely. The majority of these events are predicated on the discs evolving in an environment that is not isolated. This suggests that under many circumstances, interactions with its environment critically influence the evolution of a disc. Including this environment, either though the use of cloud-scale initial conditions or otherwise, is crucial for modelling the evolution of circumstellar discs. 

The environmentally-induced phenomena seen in our simulations are also likely to strongly affect the chemical evolution of the disc material. Previous work has shown that both gas- and ice-phase molecular abundances are highly sensitive to the evolutionary history of individual fluid parcels \citep{priestley2023a,priestley2023b,clement2023}. The dramatic differences in disc evolution we see here will therefore result in similarly-diverse disc chemistries, affecting the composition of any planetary systems which may form at later epochs \citep{oberg2021}. This will apply even to discs formed from the same parent cloud and with comparable physical properties, such as the four examples from our $\gamma_1$ simulation. These environmental effects on disc chemistry cannot be captured even by the most sophisticated MHD simulations \citep[e.g.][]{navarro2024} unless disc formation is modelled self-consistently within its cloud-scale surroundings.

\subsection{Disc Metrics are Complicated by a Dynamic Environment}

In the preceding Sections we have attempted to apply common theoretical disc models to our discs, such as the \citep{shakura_black_1973} $\alpha$ viscosity, Toomre stability criterion \citep{toomre_gravitational_1964} and $\beta$ cooling prescription \citep{gammie_nonlinear_2001}. Whilst ubiquitous in numerical studies of circumstellar discs, these models all rely on various assumptions to be valid. We have applied these criteria to our discs assuming that they are valid in their assumptions, to replicate how they are most commonly used in the study of discs. However, as we have noted throughout this study, many of these assumptions are incorrect, leading us to report anomalously high $\alpha$, $Q$ and $\beta$ values that appear to be in tension with the amount of gravitational instability and fragmentation we see. 

All the discs display evidence of GI, despite appearing Toomre-stable. Similarly the location of fragmentation events in the $\gamma_1$ discs are generally unrelated to the virial, Toomre, or $\beta$ parameters. Only the fragments in the $\gamma_{1000}$ discs are apparent in any of these metrics, but this is only because they have accumulated significant mass and are already beginning to collapse. In the $\gamma_1$ case, and the case for both  $\alpha$ and $Q$ more generally, these stability metrics seem wholly unrelated to fragmentation. 

This is a consequence of the environment around the discs and how it affects their properties and evolution. Constant, but variable, infall onto the discs affects their mass and structure, and produces discs that are almost universally massive in comparison to their host star. A different radiation environment between the $\gamma_1$ and $\gamma_{1000}$ discs affects their thermodynamic equilibrium, impacting their cooling times and the characteristic fragment mass within the disc. Interactions between discs, or between the disc and its environment, produce warps and torques that change both their structure and rotation curve. All of these external factors serve to push the discs into regimes within which the $\alpha$, $Q$ and $\beta$ parameters can no longer adequately describe them, nor would they be expected to. Therefore it is important to consider how the environment around a disc will affect the assumptions of these parameters, and whether they are applicable to the disc in such environments. It may also be pertinent to explore other methods of quantifying discs stability that are more applicable to highly time-varying environments, such as gravitational torques and both the Reynolds and gravitational stress tensors, but we leave exploration of this for a future work. 

\subsection{Limitations and Caveats}

A key limitation of this suite of simulations is the lack of full radiative transfer and protostellar radiative feedback. While we do include an approximation for the radiative emission of dust (Section \ref{sec:newPhysics}), we do not consider the effects of accretion luminosity or photoionisation. Young protostars, especially massive ones such as those formed in the $\gamma_{1000}$ clouds, can have high accretion luminosities \citep{hosokawa_evolution_2009, offner_protostellar_2011, hartmann_protostellar_2025} that will have a heating effect on their discs, often inhibiting disc fragmentation \citep{bate_importance_2009, offner_effects_2009, jones_sink_2018}. Therefore the rate of fragmentation seen in our discs may be too high, and would reduce in the presence of radiative feedback. However, it is unclear how this may interact with a higher $\gamma_{\rm SFR}$. The $\gamma_{1000}$ discs already have significantly elevated temperatures due to the external radiation field. This reduces the cooling times in the disc, which would decrease further if radiative heating were also included, encouraging further and more destructive fragmentation. Therefore we cannot predict exactly how including such physics would affect our results. 

We also do not consider the launching of outflows from our protostars, which may have an impact on the properties and evolution of the discs. Outflows remove angular momentum from discs \citep{pascucci_protostars_2023} which, depending on where the outflow is launched, may affect their properties, stability an how they fragment. Without modelling such outflows we cannot know how they would affect our discs.

Photoevaporative winds are thought to be important in dissipating circumstellar discs, producing mass loss rates up to $\rm 10^{-7} \, M_\odot \, yr^{-1}$ \citep{adams_photoevaporation_2004, williams_protoplanetary_2011, owen_theory_2012, haworth_fried_2023}. They are especially important in regions with a high ambient radiation field, as evidenced by the Orion proplyds \citep{mccaughrean_direct_1996, ricci_hubble_2008}. We do not, however, expect that photoevaporation would be important for the discs presented here. While the discs are embedded they experience accretion rates far larger than the maximum theoretical mass loss rate, meaning that any loss from photoevaporative winds would have a negligible impact on the mass of the discs. Additionally, the timescale over which these winds can dissipate the disc are much longer than the $200 \rm \, kyr$ we simulate. Only the $\gamma_{1000}$ discs may begin to experience evaporative effects after they are ejected from their natal core, but the dissipation over a few tens of kyr would be far less than the depletion the discs experience due to accretion onto the host star.

Finally, the omission of magnetic fields from these simulations is likely to affect the size and evolution of the circumstellar discs that form. Discs in simulations that include magnetic fields tend to be smaller in radius than their purely hydrodynamic counterparts, potentially better representing the observed disc population \citep{machida_effect_2011, lebreuilly_protoplanetary_2021, mayer_protostellar_2025}. However, inclusion of non-ideal MHD tends to produce disc properties closer to that of purely-hydrodynamic simulations \citep{wurster_there_2019, wurster_non-ideal_2020, mayer_protostellar_2025}. As with simulations of cloud collapse, the effect of magnetic fields is complex and potentially fundamental and it is unclear how they might interact with a changing $\gamma_{\rm SFR}$. The primary interest of this study is how the thermodynamic and structural changes to the cloud induced by $\gamma_{\rm SFR}$ affect disc formation and evolution, and we leave a consideration of magnetic fields for a future paper.

\section{Conclusions}

We have performed zoom-in simulations of molecular clouds in environments with differing interstellar radiation fields and cosmic ray ionisation rates. We have investigated the lifetimes, evolutions and fragmentation of the first circumstellar discs to form in each simulation. The conclusions of this study can be summarised as follows:

i) Changing $\gamma_{\rm SFR}$ increases the mass, temperature and surface density of the discs that form. Discs can exceed $1 \, \rm M_\odot$ in the high-$\gamma_{\rm SFR}$ clouds, and maintain temperatures in excess of $100 \rm \, K$ for most of their radius. 

ii) Protostars in high-$\gamma_{\rm SFR}$ environments accrete faster and grow more massive as a result. Their discs accrete similarly rapidly, up to and exceeding $10^{-3} \rm \, M_\odot \, yr^{-1}$. This accretion is highly time-variable for both the protostars and their discs.

iii) Discs undergo a range of varied and highly dynamical events during their lifetimes despite forming within the same parent cloud. No two discs will undergo the same evolution, owing to the differing environments around them. The properties and evolutions of the discs are strongly influenced by their surroundings, so much so that we suggest it is essential to model discs from cloud-scale initial conditions.

iv) Globally averaged stability metrics such as the Toomre and $\alpha$ viscosity are fail to adequately describe discs in realistic, chaotic environments. The assumptions these parameters rely on are generally invalid for discs in such environments, and are instead better suited to isolated, thin discs. Attempting to calculate these metrics for our discs leads to anomalously high $\alpha$ and $Q$ parameters. 

v) Disc fragmentation is common, but the mode of fragmentation differs in different $\gamma_{\rm SFR}$ environments. Discs in the low-$\gamma_{\rm SFR}$ clouds produce multiple planetary mass fragments, whereas discs in the high-$\gamma_{\rm SFR}$ clouds instead produce massive companion stars that completely disrupt the progenitor disc. 

vi) It is difficult to predict where and when fragmentation will occur in a disc based on metrics developed for isolated discs, such as the density, accretion rate, and the virial, Toomre, and $\beta$ parameters. The threshold for fragmentation appears variable with both time and environment, indicating that the process is more complex than any one model can describe. Fragmentation occurs on rapid timescales, and only the formation of massive fragments can be predicted more than 1 kyr before sink formation.

From these conclusions we support the notion that the strength of the ambient radiation field plays a critical role in the formation and evolution of discs through its effect on their thermodynamic balance, as suggested by previous studies \citep{rice_stability_2011,kratter_fragment_2011,forgan_effect_2013,rowther_short-lived_2024}. We also note that the wider environment, which in turn is affected by ambient irradiation \citep{cusack_fragmentation_2025}, is a key determinant of disc properties.

\section*{Acknowledgements}

MTC is grateful for the support of an UK Science and Technology Facilities Council (STFC) doctoral training grant. 

This research used the supercomputing facilities at Cardiff University operated by Advanced Research Computing at Cardiff (ARCCA) on behalf of the Cardiff Supercomputing Facility and the HPC Wales and Supercomputing Wales (SCW) projects. We acknowledge the support of the latter, which is part-funded by the European Regional Development Fund (ERDF) via the Welsh Government.

PCC \& FDP acknowledge the support of consolidated grant (ST/W000830/1) from the STFC. They acknowledge support from a STFC Small Award (UKRI1187): "Probing the origins of stars and life with a new approach to chemical modelling".

SCOG \& RSK acknowledge financial support from the ERC via Synergy Grant ``ECOGAL'' (project ID 855130) and from the German Excellence Strategy via the Heidelberg Cluster ``STRUCTURES'' (EXC 2181 - 390900948). In addition SCOG \& RSK are grateful for funding from the German Ministry for Economic Affairs and Climate Action in project ``MAINN'' (funding ID 50OO2206), and from DFG and ANR for project ``STARCLUSTERS'' (funding ID KL 1358/22-1).

We thank the referee for an encouraging and constructive report that improved the quality of this manuscript.


\section*{Data Availability}

The simulation snapshot data is available upon request to MTC. 


\bibliographystyle{mnras}
\bibliography{references, references-2}

\appendix

\section{Zoom Region Panel Images} \label{sec:zoomPanelImage}

\begin{figure*}
    \centering
    \includegraphics[width=0.9\textwidth]{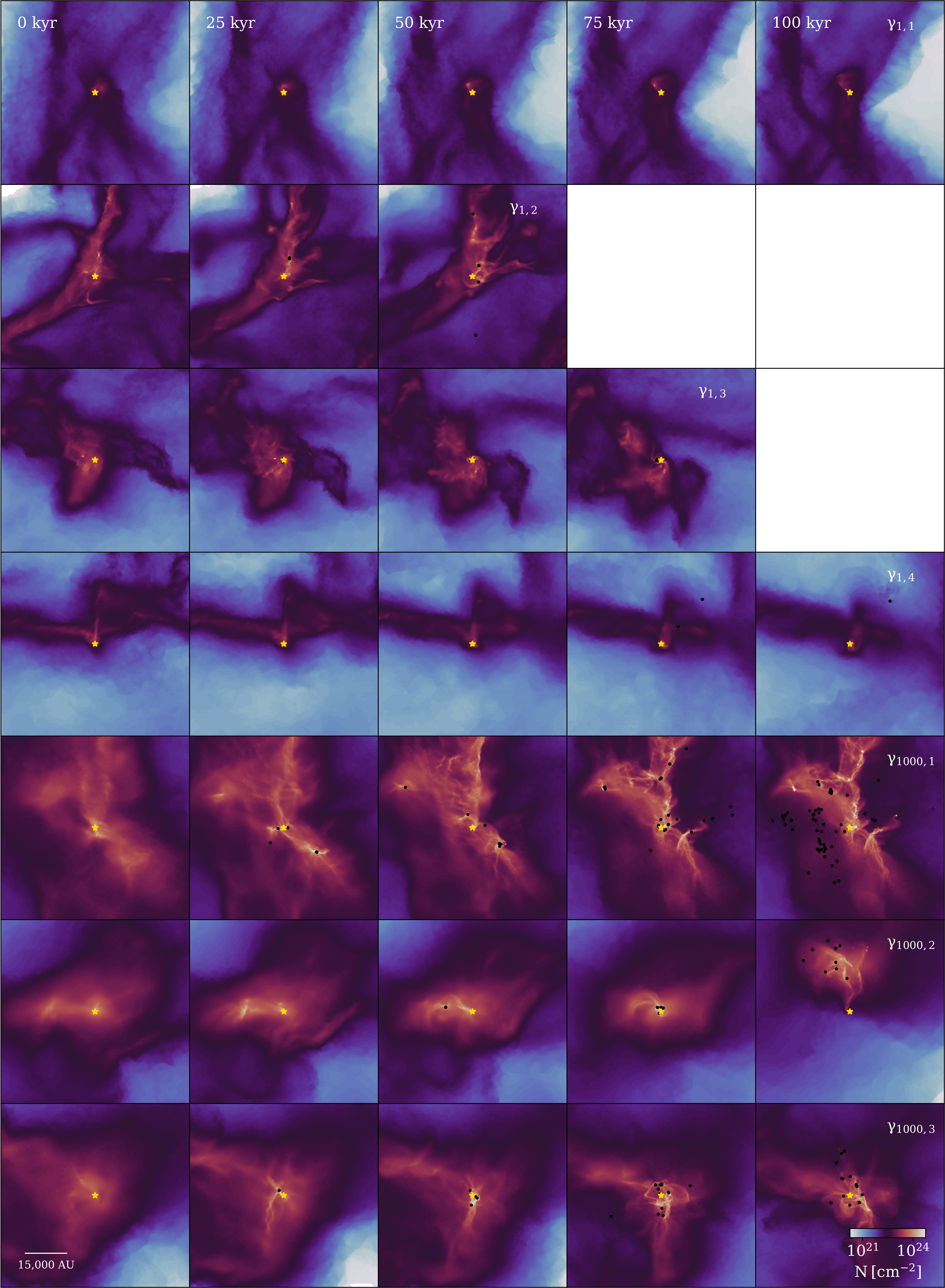}
    \caption{Column density maps of the zoom regions, centred on the first disc to form in each simulation. Each row shows the evolution of one of the zoom simulations, at intervals of 25 kyr. The time since sink formation is shown explicitly for the top row, and the final column shows the zoom simulation label for each disc. All sinks in the region are denoted with black dots, whilst the first sink is marked as a yellow star.}
    \label{fig:zoomPanel}
\end{figure*}

Figure \ref{fig:zoomPanel} shows column-density images of the zoom regions at the same time intervals as Figure \ref{fig:firstDiscsPanel}, just on a larger scale. They show how the first sink moves, or does not move, throughout the wider core. By the end of the simulation the $\gamma_1$ sinks are still firmly in the centres of their cores, whereas the $\gamma_{1000}$ sinks are not. The early ejection of the sink in $\gamma_{1000,3}$ is not immediately apparent from the Figure as it has been ejected out of the plane of the image, rather than across it as in the case of the $\gamma_{1000,2}$ simulation.

\section{Sensitivity of Disc Parameters to the Disc Identification Algorithm} \label{sec:discParameters}

\begin{figure*}
    \centering
    \includegraphics[width=\textwidth]{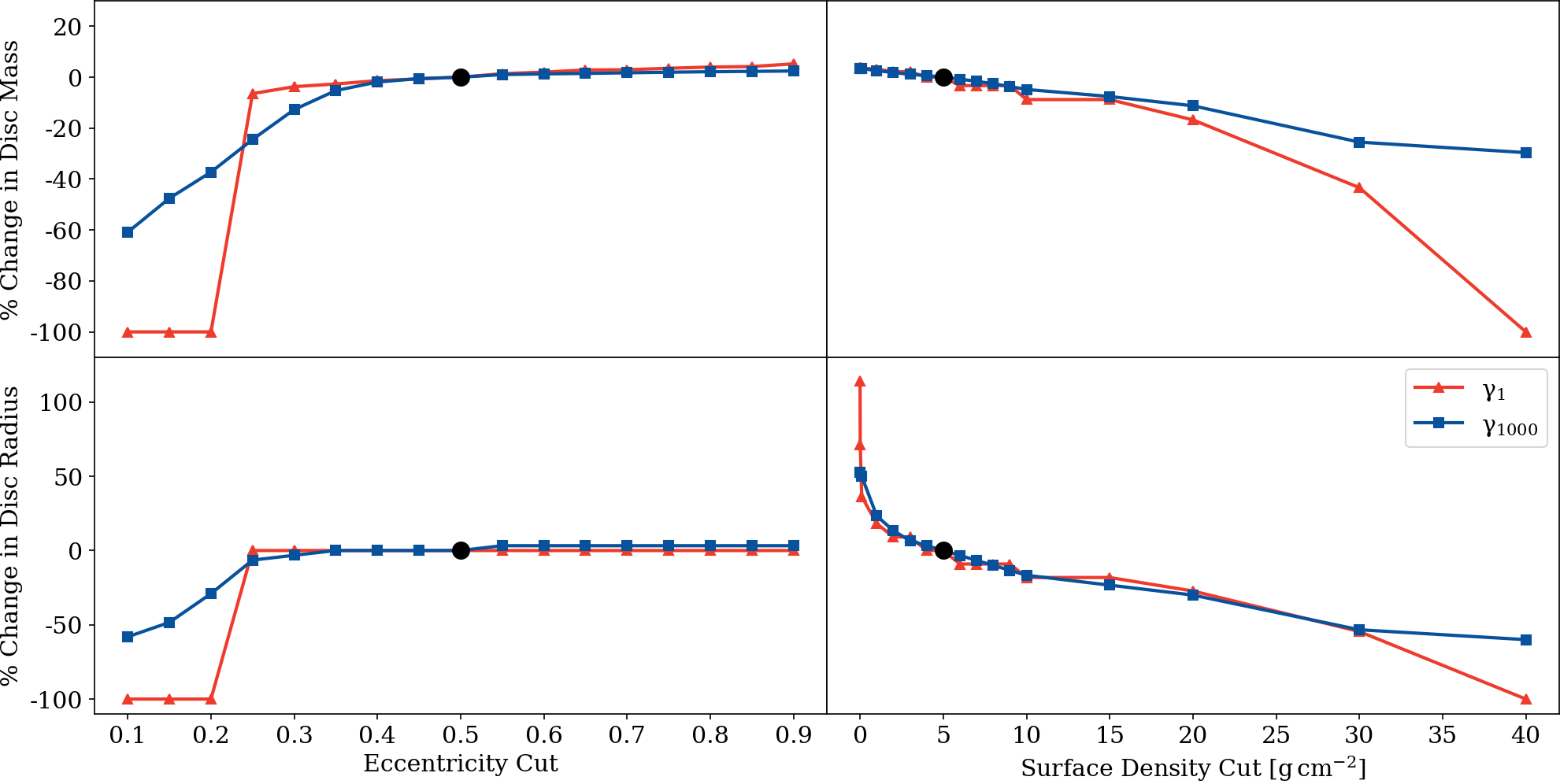}
    \caption{The variation of disc mass and radius with eccentricity and surface density cut, for two discs. The eccentricity and surface density cuts chosen for this work are denoted with a larger black circle. The $\gamma_1$ and $\gamma_{\rm 1000}$ disc values are shown with triangular and square points respectively.}
    \label{fig:discParamsPlot}
\end{figure*}

The sensitivity of the disc finding algorithm to the value of eccentricity and surface density cut is illustrated in Figure \ref{fig:discParamsPlot}. The disc mass and radius are essentially independent of eccentricity cut except for eccentricities $< 0.2$, below which mass and radii rapidly decrease. This is true for both the $\gamma_1$ and $\gamma_{\rm 1000}$ discs. Surface density is more complex, as at low surface density cuts radii rapidly increases, whereas at high cuts it decreases. The rapid increase in radius is not accompanied by an increase in mass, however, indicating that low surface density cuts include extraneous material around the disc. The chosen value for this work, $\rm 5\, g\, cm^{-2}$, lies in a midpoint between the two extremes.

\section{Calculating Map Quantities} \label{sec:mapquants}

Figures \ref{fig:fragmentationMaps1} and \ref{fig:fragmentationMaps2} show maps of the number density, local virial parameter, Toomre $Q$ and $\beta$ cooling parameter for the discs as they are about to fragment. The maps are calculated by first rotating the discs such that they are face-on. The discs are then subdivided into pixels, 1 au x 1 au in size. These pixels include cells up to 50 au above and below the plane of the disc. 

The local virial parameter map is calculated by finding the balance between gravitational potential and thermal support within the Jeans length of each pixel of the map. Within each pixel the mass weighted temperature and density are determined. These are used to calculate the Jeans length of the pixel, and all cells within its Jeans length are selected. The sum of the cells' thermal energy is calculated for $\rm E_{Support}$, and the gravitational potential energy of the ensemble of cells within the Jeans length is calculated for $\rm E_{Grav}$. These are then used to calculate the local virial parameter. Pixels with a local Jeans length > 50 au are ignored for the sake of computational efficiency. 

The Toomre parameter is calculated according to Equation \ref{eq:toomre}, where $c_{\rm s}$ is calculated using the mass-weighted average temperature of the pixel, $\kappa$ is the mass-weighted average angular velocity of cells (calculated by dividing the cells rotational velocity by their distance to the centre of the disc) within the pixel, and $\Sigma$ is the surface density of the pixel calculated by dividing its mass by the bin size.

The map of the $\beta$ parameter relies on the local cooling time. The cooling time of each cell is determined by dividing its internal energy, $u$, by the sum of all the cooling rates acting upon it, $\Sigma_i\Lambda_i$. The cooling rates are provided explicitly by the chemical network we use, allowing for an effective cooling time for each cell to be determined in a simulation that does not rely on a $\beta$-cooling prescription. For the map, the mass-weighted average cooling time of the cells within each pixel is calculated, and multiplied by the expected Keplerian angular rotation rate for the pixel to get $\beta$.

\section{Keplerian Rotation Assumption} \label{sec:kep}

\begin{figure}
    \centering
    \includegraphics[width=\columnwidth]{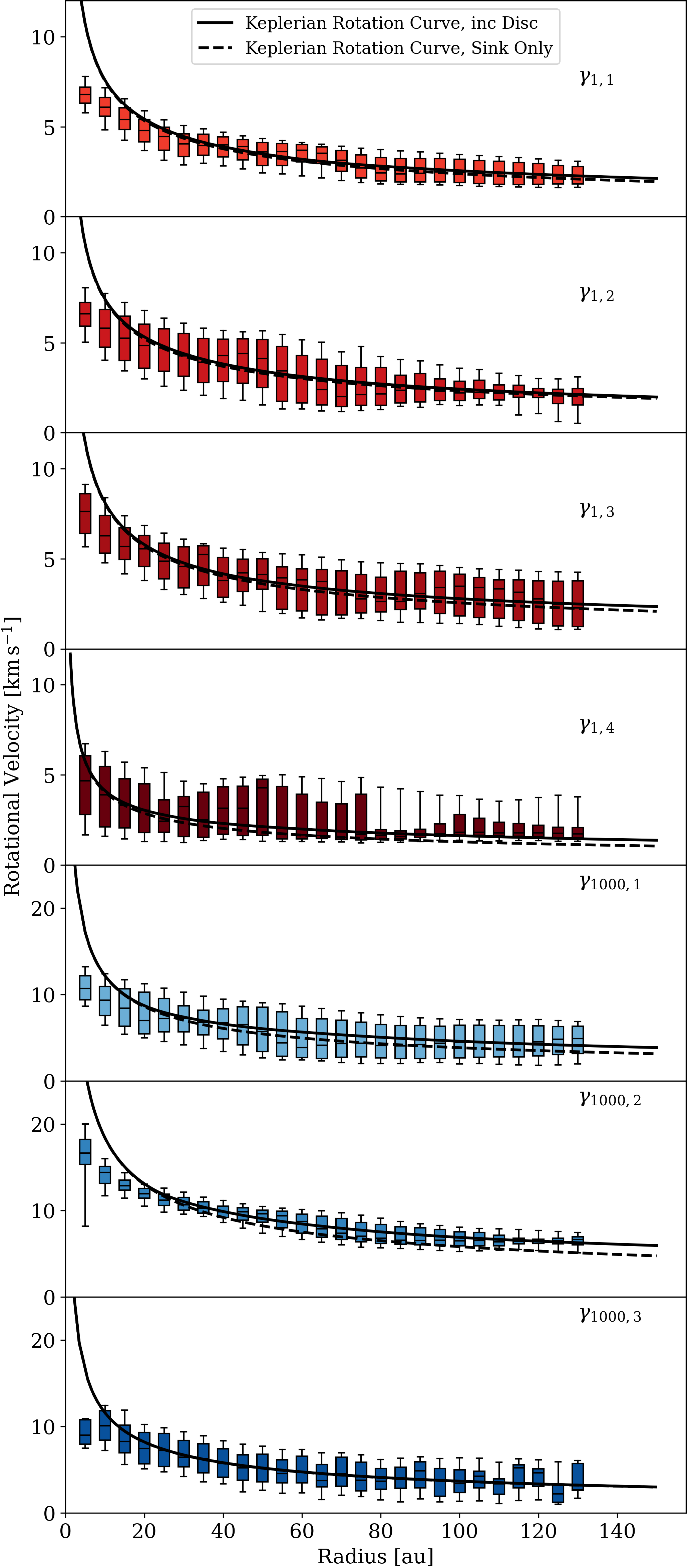}
    \caption{Rotational velocity against radius for the first discs, at a time when the discs are in a regular, unperturbed state. Box plots show the spread of rotational velocities at annuli spaced 5 au apart. The solid and dashed lines show the expected Keplerian rotation curves for the disc, including and not including the disc's mass respectively.}
    \label{fig:keplerianVelocity}
\end{figure}

In calculating the Toomre and $\alpha$ parameter we assume that the discs follow a standard Keplerian rotation curve,

\begin{equation}
    v_{\rm rot} = \sqrt{\frac{G M_*}{R}} \,,
\end{equation}

\noindent where $M_*$ is the mass of the central object. This assumption is supported by Figure \ref{fig:keplerianVelocity}, which shows examples of the rotational velocity curves of the discs. In general, the rotation curves of the discs are consistent with Keplerian rotation. 

\section{Typical Disc Densities} \label{sec:discDensities}

\begin{figure}
    \centering
    \includegraphics[width=\columnwidth]{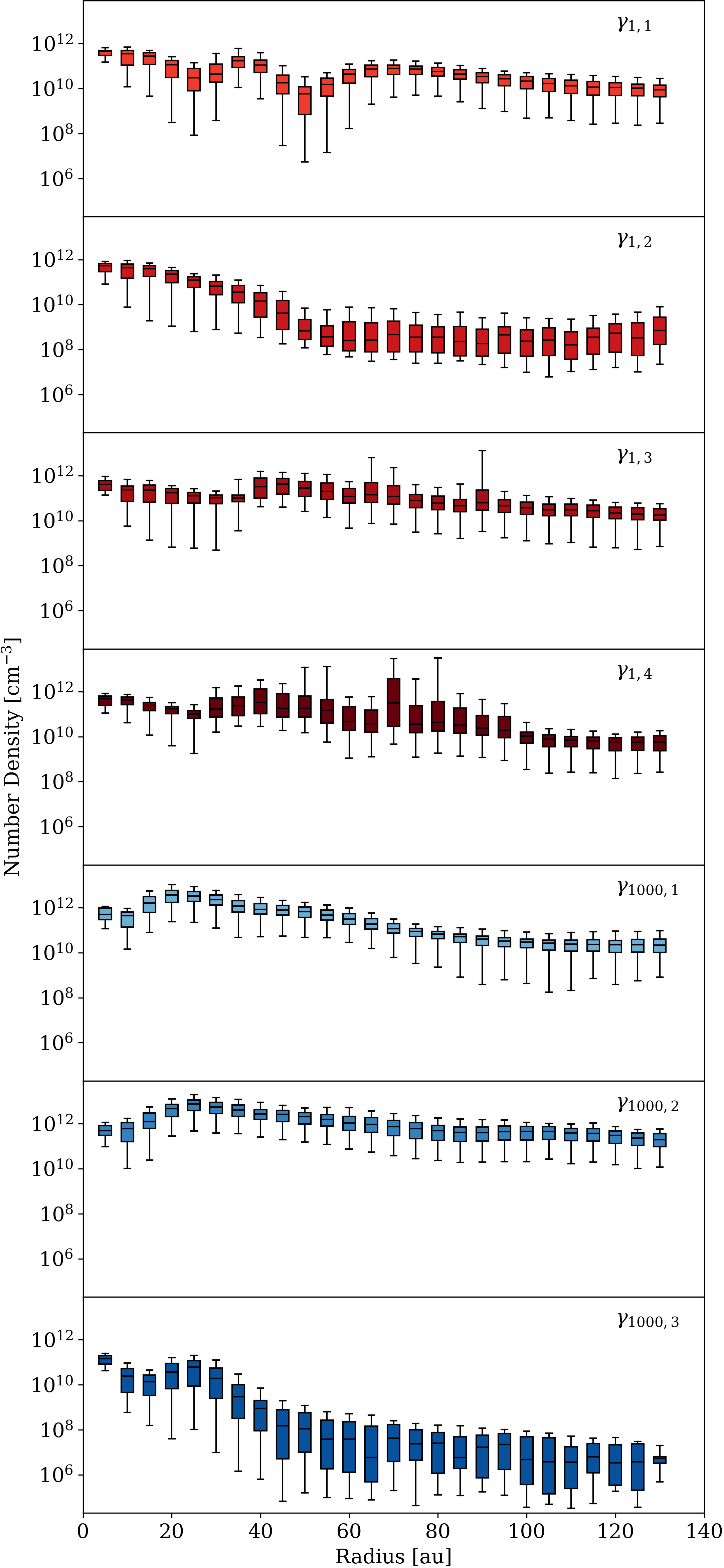}
    \caption{Number density against radius for the first discs, at a time when the discs are in a regular, unperturbed state. Box plots show the spread of number densities at annuli spaced 5 au apart.}
    \label{fig:discDensities}
\end{figure}

Figure \ref{fig:discDensities} illustrates the typical number densities of the discs across their radii, at a time when they are in an unperturbed state, for reference. 

\bsp	
\label{lastpage}
\end{document}